\documentclass[aps,prb,twocolumn,floats,floatfix,groupedaddress]{revtex4-2}
\usepackage{amsmath}
\usepackage{amssymb}
\usepackage{graphicx}
\usepackage{siunitx}
\usepackage{longtable}
\usepackage{multirow}
\usepackage{upgreek}
\begin{document}

%\preprint{}

\title{Caustic spin wave beams in soft, thin films: properties and classification}


%\author{}
%\thanks{}
%\affiliation{}

\author{Alexis Wartelle}
\email[]{alexis.wartelle@ens-lyon.org}
%\thanks{}
\altaffiliation[Present address: ]{Universit\'{e} Grenoble Alpes, CNRS, Grenoble INP, SIMaP, 38000 Grenoble, France}

\author{Franz Vilsmeier}

\author{Takuya Taniguchi}

\author{Christian H. Back}

\affiliation{Fakult\"{a}t fur Physik, Technische Universit\"{a}t M\"{u}nchen, Garching, Germany}


\date{\today}

\begin{abstract}
In the context of wave propagation, caustics are usually defined as the envelope of a finite-extent wavefront; folds and cusps in a caustic result in enhanced wave amplitudes. Here, we tackle a related phenomenon, namely the existence of well-defined beams originating solely from the geometric properties of the corresponding dispersion relation. This directional emission, termed caustic beam, is enabled by a stationary group velocity direction, and has been observed first in the case of phonons. We propose an overview of this ``focusing'' effect in the context of spin waves excited in soft, thin ferromagnetic films. Based on an analytical dispersion relation, we provide tools for a systematic survey of caustic spin wave beams. Our theoretical approach is validated by time-resolved microscopy experiments using the magneto-optical Kerr effect. Then, we identify two cases of particular interest both from fundamental and applicative perspectives. Indeed, both of them enable broadband excitations (in terms of wave vectors) to result in narrowband beams of low divergence.
\end{abstract}


\maketitle

\section{Introduction}

The collective motion of magnetic moments in a materials, referred to as spin waves, has shown remarkable properties from a fundamental perspective. Examples range from anisotropic dispersion in thin films \cite{Yu2020_ReviewMagnonics}, relevant for the field of magnonics, to Bose-Einstein condensation of magnons \cite{Divinskiy2021_MagnonBEC}, through restricted-relativity-like bounded domain wall velocities \cite{Bouzidi1990_BlochWallMotionWithMagnonicLimit}. Applications of magnetization dynamics also abound, starting with the infinite-wavelength ferromagnetic resonance (FMR) \cite{Stancil2009} and going all the way towards sub-micrometer wavelengths, which are currently viewed as promising alternative information carriers in the fields of magnonics \cite{Chumak2019_FundamMagnonics_swCaustics}. In addition to the absence of Joule heating and the potential device downscaling (using small wavelengths), spin wave interference is an appealing prospect \cite{Petti2022_ReviewMagnonicsUsingSpinTextures} as it allows logic operations through the design of the propagation lines.

Several experimental techniques are readily available for the study of spin waves \cite{Yu2020_ReviewMagnonics}, especially in the case of thin films or patterned elements thereof. Among them, micro-/phase-resolved Brillouin Light Scattering (BLS) \cite{Seo2021_SWexcitationInPyRectangles}, Time-Resolved Magneto-Optical Kerr Effect (TR-MOKE) microscopy \cite{Au2011_TRMOKE_SpinWaveObservation}, and time-resolved Scanning Transmission X-ray Microscopy (TR-STXM) with magnetic sensitivity through X-ray Magnetic Circular Dichroism (XMCD \cite{Stoehr2006}) \cite{Sluka2019_1Dand2DSWsInAFcoupledVortexLayers} have demonstrated outstanding imaging capabilities. Nevertheless, the usually very small amplitudes of magnetization precession associated to spin waves as well as their attenuation lengths (typically on the micrometer scale) pose a significant challenge both for fundamental investigations and for applications.

To be of practical use, spin waves must be harnessed via a power-efficient strategy: some approaches like Winter's magnons rely on channeling along domain walls \cite{Aliev2011_WintersMagnonsInPyDots}, others rely on careful control of spin wave scattering \cite{Golebiewski2020_TalbotEffectForForwardVolumeSWs}.  Another possibility would take advantage of caustic spin wave beams (CSWBs), \emph{i.e.}  spin wave beams of well-defined propagation direction, narrow angular width and higher power compared to \emph{e.g.} Damon-Eshbach-type \cite{Damon_Eshbach1961_SWmodesInThinFilms__RefPaper} spin waves. Furthermore, caustics in soft, thin ferromagnetic films can be very different from the well-known acoustical or optical caustics, which originate from inhomogeneous media \cite{Kravtsov1990_GeometricalOpticsOfInhomogeneousMedia,Kravtsov1993_CausticsCatastrophesAndWaveFields,Poston1996_CatastropheTheoryAndItsApplications}, : here, spin wave caustics can arise in perfectly homogeneous films in broad ranges of conditions solely because of sufficient anisotropies in their dispersion relation. The latter indeed allows the direction of the group velocity to be stationary around some wave vectors, leading to well-defined directions of wave propagation associated to significantly stronger emission. In the context of phonon propagation, such phenomena have been referred to as ``focussing'' \cite{Northrop1980_BallisticPhononImagingInGe}, and they have been observed and investigated since 1969 \cite{Taylor1969_FirstObservationOfPhononCaustics,Maris1971_PhononFocusingDiscussedAsAnisotropicDR,Northrop1980_BallisticPhononImagingInGe,Every1987_PhononCausticsInQuartz,Maris1986_ReviewOnPhononFocusing}.

By contrast, caustics in ferromagnetic films were reported for the first time \emph{ca.} 30 years later \cite{Buettner2000_LinAndNonLinSWscatteringInYIGbyBLS}. There has been quite a few reports since then \cite{Veerakumar2006_CausticsInYIG_perhapsFirstPaper,Khomeriki2004_CausticMagnetostaticSWbeamsWithNLschroedingerEq,Demidov2009_SWfromStripeWaveguideIntoFilm_CausticBeams,Schneider2010_CausticsInYIG,Kostylev2011_CausticSWbeamsFromDecayingSWwavepacket,Sebastian2013_CausticSWbeamsFromEdgeModeInHeuslerAlloy,Kim2016_PredictionOfSWcausticsInPerpFilmsWithDMI,Heussner2017_SWsplitter_wCausticBeams_DirectionCtrledByOerstedField,Muralidhar2020_SWcausticsFromFSlaserPulseOnPy}  but, to the best of our knowledge, there exists to date no systematic survey of the properties of spin wave caustics, not even focusing on a certain type of systems \emph{e.g.} ultrathin films with perpendicular anisotropy, or soft thin films. In this work, we restrict ourselves to the latter and give an overview of caustics in soft thin films, as well as tools to further investigate them. Moreover, we highlight two special cases which seem particularly appealing notably for application in magnonics.

\section{\label{sec_model}Model}

\subsection{General considerations}

Our starting point is the model derived by Kalinikos and Slavin \cite{Kalinikos_Slavin1986_DipoleExchangeSWmodesInThinFilms_MixedBoundConditions} for spin waves in soft ferromagnetic thin films. These excitations correspond to a time- and space-dependent magnetization $\overrightarrow{M}(\vec{r},t)$, yet its norm $M_\mathrm{s}=||\overrightarrow{M}(\vec{r},t)||$ the spontaneous magnetization is uniform. As a result, it is simpler to consider the reduced magnetization $\overrightarrow{m}(\vec{r},t)=\overrightarrow{M}(\vec{r},t)/M_\mathrm{s}$ with norm 1. We focus on the linear regime \emph{i.e.} the deviation $\delta \overrightarrow{m}(\vec{r},t)=\overrightarrow{m}(\vec{r},t)-\overrightarrow{m}_0(\vec{r},t)$ from the equilibrium magnetization (when no excitation is applied) $\overrightarrow{m_0}$ is such that $||\delta \overrightarrow{m}||\ll 1$. Under the assumption of negligible mode mixing and of a perfectly isotropic ferromagnetic material, one may write the dispersion relation of a thin film as:

%\begin{equation}
%\begin{split}\label{eq_Kalinikos_Slavin}
%\omega^2	=& \Big[\gamma_0 H_\mathrm{a}+ \frac{2A\gamma_0}{\mu_0 M_\mathrm{s}}k^2\Big]\Big\{\gamma_0 \Big[M_\mathrm{s}+H_\mathrm{a}\Big]+ \frac{2A\gamma_0}{\mu_0 M_\mathrm{s}}k^2\Big\}-\gamma_0^2 M_\mathrm{s}^2\cdot\xi(kd)\Big[1-\xi(kd)+\frac{H_\mathrm{a}}{M_\mathrm{s}}+ \frac{2A\gamma_0}{\mu_0 M_\mathrm{s}^2}k^2\Big]\cos^2{\varphi} \\
%			& +\gamma_0^2 M_\mathrm{s}^2\cdot\xi(kd)\cdot[1-\xi(kd)]
%\end{split}\end{equation}
\begin{eqnarray}
\omega^2	&=& \Big[\gamma_0 H_\mathrm{a}+ \frac{2A\gamma_0}{\mu_0 M_\mathrm{s}}k^2\Big]\Big[\gamma_0 \Big(M_\mathrm{s}+H_\mathrm{a}\Big)+ \frac{2A\gamma_0}{\mu_0 M_\mathrm{s}}k^2\Big] \nonumber\\
			&& -\gamma_0^2 M_\mathrm{s}^2\cdot\xi(kd)\Big[1-\xi(kd)+\frac{H_\mathrm{a}}{M_\mathrm{s}}+ \frac{2A\gamma_0}{\mu_0 M_\mathrm{s}^2}k^2\Big]\cos^2{\varphi} \nonumber\\
			&& +\gamma_0^2 M_\mathrm{s}^2\cdot\xi(kd)\cdot[1-\xi(kd)]
\label{eq_Kalinikos_Slavin}\end{eqnarray} \\ where $\omega$ is the spin wave angular frequency, $\gamma_0=\mu_0|\gamma|$ with $\gamma=q_\mathrm{e}/(2m_\mathrm{e})$ the electron's gyromagnetic ratio ($q_\mathrm{e}=-e$ and $m_\mathrm{e}$ being the electron's charge and mass, respectively) and $\mu_0$ the permeability of vacuum, $A$ is the micromagnetic exchange constant for the soft ferromagnetic material of interest, $M_\mathrm{s}$ its spontaneous magnetization, $k$ the spin wave's wavenumber corresponding to its wave vector $\overrightarrow{k}$, $H_\mathrm{a}=||\overrightarrow{H_\mathrm{a}}||$ the strength of the externally applied magnetic field $\overrightarrow{H_\mathrm{a}}$ from which $\varphi=\mbox{angle}\Big(\overrightarrow{H_\mathrm{a}},\overrightarrow{k}\Big)$ the wavefront angle is defined, $d$ the film thickness, and $\xi$ is the function whose values are defined as:

\begin{equation}
\xi(u)=1-\frac{1-e^{-u}}{|u|}.
\end{equation}

\begin{figure}[h!]
\includegraphics[width=7.599cm]{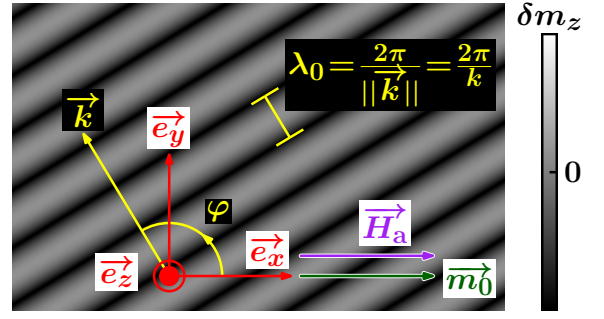}
\caption{\label{fig_orientations}Schematic representation of a spin plane wave propagating in a soft thin film. The grey scale codes the local perpendicular component of the dynamic component of magnetization, $\delta m_z$.}
\end{figure}

As a consequence of the ferromagnetic material's softness, in the absence of excitation, the equilibrium magnetization configuration in our thin film is the single-domain state, with a corresponding reduced magnetization $\overrightarrow{m}_0$ %=\overrightarrow{M}_0/M_\mathrm{s}$ 
exactly along the applied field. The orientations of $\overrightarrow{m}_0$, $\overrightarrow{H_\mathrm{a}}$, and $\overrightarrow{k}$ are illustrated in Fig. \ref{fig_orientations}, which also highlights the natural wavelength $\lambda_0=2\pi/||\overrightarrow{k}||$ of the spin wave as well as the unit vectors $\overrightarrow{e_x}$, $\overrightarrow{e_y}$ and $\overrightarrow{e_z}$.

Here, we focus on spin waves with no amplitude node across the film thickness, \emph{i.e.} we do not consider perpendicular standing spin waves (PSSWs). However, we do note that the latter may play a role in experiments performed on sufficiently thick films where a realistic antenna for instance could excite them due to its inhomogeneous magnetic field. 

We introduce the following quantities: the Larmor angular frequencies associated to magnetization $\omega_\mathrm{M}=\gamma_0 M_\mathrm{s}$ and to the applied magnetic field $\omega_\mathrm{H}=\gamma_0 H_\mathrm{a}$, the material's dipolar-exchange length $l_\mathrm{ex}=\sqrt{2A/(\mu_0M_\mathrm{s}^2)}$. We then rewrite the equation as:

\begin{eqnarray}
\frac{\omega^2}{\omega_{\mathrm{M}}^2}	&=& \left(\frac{\omega_\mathrm{H}}{\omega_\mathrm{M}}+ l_\mathrm{ex}^2k^2\right)\left(1+\frac{\omega_\mathrm{H}}{\omega_\mathrm{M}}+l_\mathrm{ex}^2 k^2\right) \nonumber\\
&& -\xi(kd)\Big[1-\xi(kd)+\frac{\omega_\mathrm{H}}{\omega_\mathrm{M}}+ l_\mathrm{ex}^2 k^2\Big]\cos^2{\varphi} \nonumber\\
&&+\xi(kd)\Big[1-\xi(kd)\Big]
\end{eqnarray}

Introducing the reduced frequency $\nu=\omega/\omega_\mathrm{M}$ and applied field $h=\omega_\mathrm{H}/\omega_\mathrm{M}=H_\mathrm{a}/M_\mathrm{s}$, and normalizing both the dipolar-exchange length and wavenumber to the film thickness $d$ using $\eta=l_\mathrm{ex}/d$ and $\tilde{k}=kd$, we arrive at:

\begin{eqnarray}
\nu^2 &=& \Big(h+ \eta^2\tilde{k}^2\Big)\Big(1+h+ \eta^2\tilde{k}^2\Big) \nonumber\\
&&-\xi(\tilde{k})\Big[1-\xi(\tilde{k})+h+ \eta^2\tilde{k}^2\Big]\cos^2{\varphi} \nonumber\\
&&+\xi(\tilde{k})\Big[1-\xi(\tilde{k})\Big]
\label{eq_DR}\end{eqnarray}

With this, it is clear that any given experiment of spin wave excitation corresponds to a specific value of the dimensionless triplet ($\eta$, $\nu$, $h$). In other words: they are the only independent parameters within this model.

For a value of ($\eta$, $\nu$, $h$), the solution to \eqref{eq_DR} is the possibly empty set of accessible dimensionless wave vectors $\overrightarrow{k}d$. The existence and properties of spin wave caustics depend on the geometrical characteristics of this set, which is why we are first going to review several of its general properties.

Keeping in mind that we focus on applied fields below the ferromagnetic resonance field at the excitation frequency, we actually always have a non-empty solution, which is usually a closed curve winding around the origin in wave-vector space. This is the so-called slowness curve, in reference to the fact that at fixed frequency $k\propto 1/||\overrightarrow{v_\mathrm{p}}||$ where $\overrightarrow{v_\mathrm{p}}$ is the phase velocity \cite{Lax1980_CausticEnhancementEvaluationAndDetectorShapeEffects}, oriented of course along the wave vector. Considering the parity of the cosine function and its antisymmetry for the reflection $\varphi\rightarrow \pi-\varphi$, we may restrict our analysis to only the quadrant $\varphi\in[0,\pi/2]$ and deduce the others using mirror symmetries. 

One can also parametrize the slowness curve using a curvilinear abscissa: we define it to be zero for the lowest dimensionless wavenumber $\tilde{k}_\mathrm{min}$ at $\varphi=\pi/2$. One can indeed show that the reduced wavenumber solving Eq. \eqref{eq_DR} at $\varphi=\pi/2$ (resp. $0$) is minimum (resp. maximum) on the quadrant $\varphi\in[0,\pi/2]$. Thus, at the largest dimensionless wavenumber $\tilde{k}_\mathrm{max}=\tilde{k}(\varphi=0)$, the corresponding curvilinear abscissa $s_\mathrm{M}$ corresponds to the length of the slowness curve in the quadrant $\varphi\in[0,\pi/2]$ \emph{i.e.} one fourth of the whole length of this curve.
 
Another important geometrical aspect of the slowness curve that is central to the present work is the local normal to it. Considering its definition as a constant-frequency intercept of the dispersion relation in wave-vector space, by nature, the frequency gradient $\overrightarrow{\nabla}_{\hspace{-0.15cm}\overrightarrow{k}}\omega$ is perpendicular to the slowness curve. As a result, the direction of the group velocity of spin waves $\overrightarrow{v_\mathrm{g}}=\overrightarrow{\nabla}_{\hspace{-0.15cm}\overrightarrow{k}}\omega$ can be directly read from the direction of the local normal to the slowness curve. In our notations, we point out that:

\begin{equation*}
\overrightarrow{\nabla}_{\hspace{-0.15cm}\overrightarrow{k}}\omega\equiv\sum_{\beta=x,y,z}\frac{\partial\omega}{\partial k_\beta}\cdot\overrightarrow{e_\beta}\qquad\mbox{where}\quad k_\beta=\overrightarrow{k}\cdot\overrightarrow{e_\beta}.
\end{equation*}

\begin{figure}%[h!]
\includegraphics[width=7.0cm]{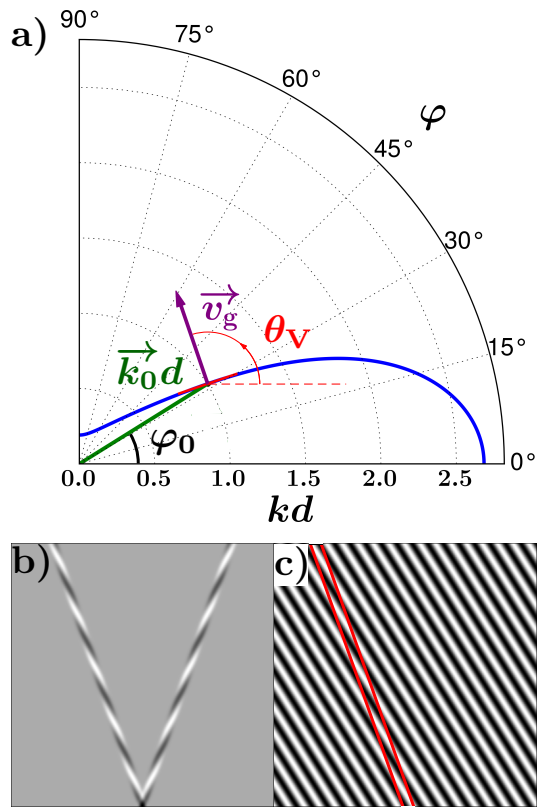}
\caption{\label{fig_slownesscurve}a) Exemplary slowness curve for $(\nu, h, \eta)=(0.2873, 10^{-20}, 0.15)$. As can be clearly seen in the polar plot of $kd=\tilde{k}(\varphi)$, the direction ($\varphi_0=$\ang{32.00}) of the phase velocity $\protect\overrightarrow{v_\mathrm{p}}$ and that ($\theta_\mathrm{V}=$\ang{108.9}) of the group velocity $\protect\overrightarrow{v_\mathrm{g}}$ at the point $\protect\overrightarrow{k_0}d$ are very different. b) Radiation pattern ($\delta m_z$ is grey-coded) of a hypothetical source exciting only wavenumbers very close to $||\protect\overrightarrow{k_0}||d$. c) Plane wave corresponding to the carrier wave vector $\protect\overrightarrow{k_0}d$  (red lines are guides to the eye).}
\end{figure}

 In the following, we will use the angle $\theta_\mathrm{V}=\mathrm{angle}(\overrightarrow{H_\mathrm{a}},\overrightarrow{v_\mathrm{g}})$. We point out that in the present case, phase and group velocities need not be collinear: on the contrary, there can be differences between $\theta_\mathrm{V}$ and $\varphi$ much larger than in cases of light propagation through anisotropic media \footnote{Even for markedly anisotropic dielectrics the differences in refractive indices lead to maximal angles between phase and group velocities below \ang{10} \cite{HandbookOfOpticsVol2}.}. Fig.~\ref{fig_slownesscurve} illustrates this on the example of a slowness curve reconstructed for a vanishing reduced applied field.

%%\begin{figure*}%[h!]
%%\includegraphics[width=16.79cm]{fig_Slownesscurve.pdf}
%%%\includegraphics[width=8.599cm]{fig_Slownesscurve.pdf}
%%\caption{\label{fig_slownesscurve}a) Exemplary slowness curve for $(\nu, h, \eta)=(0.2873, 10^{-20}, 0.15)$. As can be clearly seen in the polar plot of $kd=\tilde{k}(\varphi)$, the direction ($\varphi_0=$\ang{32.00}) of the phase velocity $\protect\overrightarrow{v_\mathrm{p}}$ and that ($\theta_\mathrm{V}=$\ang{108.9}) of the group velocity $\protect\overrightarrow{v_\mathrm{g}}$ at the point $\protect\overrightarrow{k_0}d$ are very different. b) Radiation pattern ($\delta m_z$ is grey-coded) of a hypothetical source exciting only wavenumbers very close to $||\protect\overrightarrow{k_0}||d$. c) Plane wave corresponding to the carrier wave vector $\protect\overrightarrow{k_0}d$  (red lines are guides to the eye).}
%\end{figure*}

\subsection{\label{sec_distinctiveFeats}Distinctive features of dispersion relation caustics}

Typically, caustics in inhomogeneous media occur when a wavefront folds onto itself; in this situation, there exists a surface (or a line in 2D wave propagation) such that across it the number of rays passing through a point in space changes by an even number \cite{Kravtsov1993_CausticsCatastrophesAndWaveFields,Poston1996_CatastropheTheoryAndItsApplications}: this is the caustic. Equivalently, it can be viewed as the set of the local extrema of positions on the ray bundle on the wavefront, for all the wavefronts along the wave propagation. It is this extremal nature that grants these caustics large and localized intensities compared to other points on the ray bundle. In a geometrical optics approach, the intensity diverges as an initially finite-sized portion of the wavefront shrinks to a vanishing area \cite{Kravtsov1993_CausticsCatastrophesAndWaveFields}. A wave optics treatment however reveals that the intensity remains finite due to interferences: illumination profiles across caustics can in principle be determined by taking into account the variations of phase as a function of distance to the caustic \cite{Kravtsov1990_GeometricalOpticsOfInhomogeneousMedia}.

Such an approach has been used by Schneider \emph{et al.} \cite{Schneider2010_CausticsInYIG} for spin wave caustics excited by the scattering of a spin wave travelling in a waveguide terminating into a full permalloy (Ni$_{80}$Fe$_{20}$) film. However, this is a very different situation compared to the above. Indeed, the wavefront does not fold onto itself due to spatial variations of medium properties, rather, its extent is determined almost exclusively (owing to the sub-wavelength source size) by the characteristics of spin wave propagation. The latter are determined by the anisotropic spin wave dispersion relation, which allows caustics to form thanks to the possibility of stationary group velocity direction \emph{i.e.} a beam with a well-defined propagation direction yet comprising a range of wave vectors in the vicinity of a carrier. More precisely, caustics correspond to local extrema of the group velocity direction; in other words, a caustic spin wave beam implies the existence of a caustic point $\tilde{k}_\mathrm{c}$ on the slowness curve such that:

\begin{equation}
\left.\frac{\mathrm{d}\theta_\mathrm{V}}{\mathrm{d}\tilde{k}}\right|_{\tilde{k}_\mathrm{c}}=0.
\end{equation}

The CSWB has then a carrier wavenumber $\tilde{k}_\mathrm{c}$, corresponding to a central wavefront angle $\varphi_\mathrm{c}=\varphi(\tilde{k}_\mathrm{c})$ and a beam direction $\theta_\mathrm{V,c}=\theta_\mathrm{V}(\tilde{k}_\mathrm{c})$.

Coming back to the wavefront extent, rays from wave vectors not close enough to the carrier cannot play a role in the caustic wave amplitude simply because of differences in propagation direction. More specifically, the experimental data presented by Schneider \emph{et al.} suggests that beam divergences of \ang{2} or less can be obtained. Thus, there seems to be a contradiction between the cubic dispersion which is assumed to define the beam profile and the measurements. The question of the CSWB's profile goes however beyond the scope of this work. Nevertheless, it is clear from the low beam divergences observed in many experimental reports \cite{Bertelli2020_MultipleSWcausticsFromStriplineCorners,Kostylev2011_CausticSWbeamsFromDecayingSWwavepacket,Veerakumar2006_CausticsInYIG_perhapsFirstPaper} that only small, almost straight parts of the slowness curve must contribute to CSWB.

In fact, integrating the contribution of wave vectors all the way to infinity as done in \cite{Schneider2010_CausticsInYIG} neglects the geometric impossibility for them to create waves travelling from the point source to a far-away point \emph{on the caustic}. To put it differently: for geometrical reasons, caustics originating solely from anisotropies in the dispersion relation and excited by a point-like source naturally restrict the range of relevant wave vectors, in contrast to the case of caustics due to inhomogeneities in the propagation medium.

We wish to emphasize the above by reminding that in most cases \cite{Felsen1994_RadiationAndScatteringOfWaves,Kravtsov1993_CausticsCatastrophesAndWaveFields}, caustics are treated on the basis of wave propagation in an isotropic or weakly anisotropic medium. One consequence is the fact that the flow of power, \emph{i.e.} the group velocity, is along the wave vector or close to parallel to it \cite{Kravtsov1990_GeometricalOpticsOfInhomogeneousMedia}. While this remains a reasonable approximation for slightly anisotropic media (as in usual crystal optics), in the case of perfectly soft but fully polarized thin ferromagnetic films this collinearity may break down dramatically, as was illustrated in Fig. \ref{fig_slownesscurve}. 
%A simple explanation is that spin waves in such media do not obey a Helmholtz-type equation (notably due to the coupling of oscillating magnetization components). 
Therefore, even small changes in wave vector may result in drastic changes in group velocity direction. By contrast, large changes in wave vectors do not necessarily lead to strong variations in the \emph{apparent} wavelength $\lambda$ which we define as:

\begin{equation}
\lambda=\frac{2\pi\cdot ||\overrightarrow{v_\mathrm{g}}||}{\overrightarrow{k}\cdot\overrightarrow{v_\mathrm{g}}}=\frac{2\pi}{\overrightarrow{k}\cdot \overrightarrow{e_\mathrm{g}}}=\frac{\lambda_0(\varphi)}{\cos{(\theta_\mathrm{V}-\varphi)}},
\end{equation} \\ where we have introduced $\overrightarrow{e_\mathrm{g}}$ as a unit vector along the group velocity. The apparent wavelength is simply the spatial period measured along the beam direction. Since large differences $\theta_\mathrm{V}-\varphi$ can easily be obtained (\emph{cf.} Fig. \ref{fig_slownesscurve}, where $\cos{(\theta_\mathrm{V}-\varphi_0)}\simeq 0.227$), and more importantly since the projection $\tilde{k}(\varphi)\cos{(\theta_\mathrm{V}-\varphi)}$ may remain almost constant over significant portions of the slowness curve, one should consider notions such as propagation-induced phase or spectral breadth \footnote{We define the spectral breadth of a non-monochromatic beam with well-defined propagation direction and negligible divergence as $\Delta\lambda/\lambda_\mathrm{car.}$, where $\lambda_\mathrm{car.}$ is the carrier's mean apparent wavelength, and $\Delta\lambda$ is the beam's standard deviation in apparent wavelength.} of a spin wave beam carefully.

\section{\label{sec_results}Results and discussion}

\subsection{\label{sec_thick_films}Limit of model applicability: thick films}

\begin{figure}[h!]
\includegraphics[width=7.599cm]{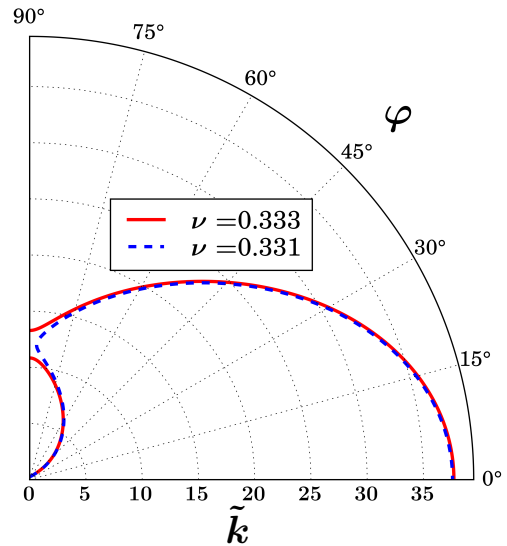}
\caption{\label{fig_connectedComp_thickfilms}Slowness curves for $\eta=0.015$, $h=10^{-20}$, and $\nu=0.331$ (dashed blue line) resp. $\nu=0.333$ (full red line).}
\end{figure}

We start by providing an example of situation where the model we use cannot be fully trusted, so as to highlight its limitations. In Fig.~\ref{fig_connectedComp_thickfilms} we show a case where the reconstructed slowness curve splits into two separate connected components above a certain threshold frequency.

%\vspace{-0.9\baselineskip}

Such a behaviour has been described by Kreisel \emph{et al.} \cite{Kreisel2009_FlawsOfsualSWdispersionInDipExchRegime}: the model chosen for spin wave dispersion predicts a local maximum in the $\omega(k,\varphi=\pi/2)$ \emph{vs.} wavenumber curve, but this extremum is not reproduced by a formal approach not based on the thin-film approximation \cite{Harms2022_DipoleExchangeSpinWaveTheory}, and designed to tackle the dipole-exchange regime. The maximum's presence leads to an additional pair of solutions in terms of wavenumber in a certain frequency range, corresponding to a splitting of the slowness curve into two separate components. 

Clearly, results obtained within our approach about caustics deep in the dipole-exchange regime are not trustworthy. Empirically, we see the slowness curve splitting into separate components for values of $\eta$ up to \emph{ca.} $0.075$; for the sake of comparison, the thinnest films investigated by Kreisel \emph{et al.} feature $\eta< 0.035$ according to literature data on yttrium iron garnet (YIG) \cite{Klingler2014_fmrMeasurementOfYIGexchangeCst}. Nevertheless, the absence of this splitting is no proof that the reconstructed slowness curve is accurate, and we shall remain cautious in discussing results concerning CSWBs with wavenumbers in the dipole-exchange regime.  Finally, we note that promising theoretical developments such as the dipole-exchange dispersion relations recently derived by Harms and Duine \cite{Harms2022_DipoleExchangeSpinWaveTheory} could eventually allow a more accurate treatment of caustics in the dipole-exchange regime.

\subsection{\label{sec_general_features}General features}

\begin{figure*}[!h] 
\includegraphics[width=17.59cm]{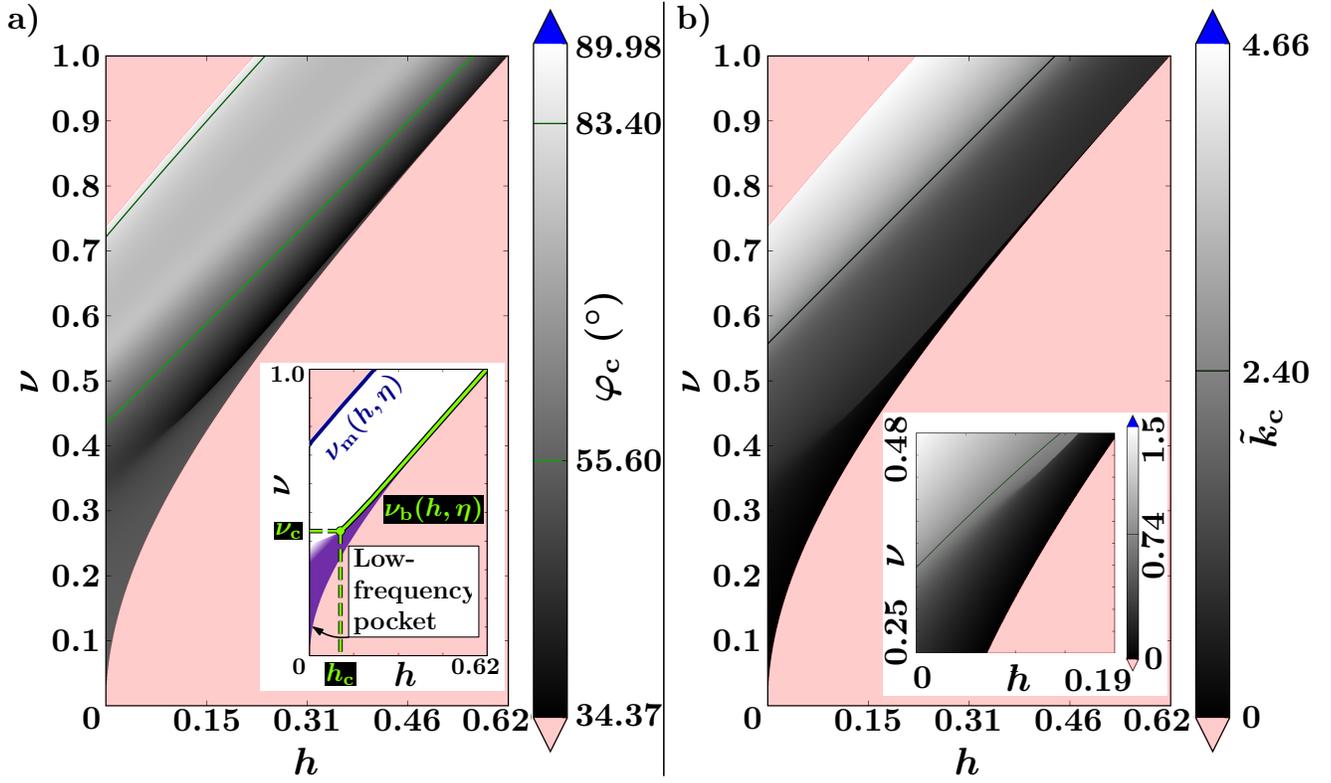}
\caption{\label{fig_ex_fieldfreq_maps}Frequency and field maps for a value of $\eta=0.12$. For high enough fields, a sharp upturn in both properties can be seen for reduced frequencies above \emph{ca.} 0.42. We remind the reader that fields above ferromagnetic resonance are not considered. Only few level curves are displayed for the sake of clarity. a) Caustic wavefront angle $\varphi_\mathrm{c}$, with a schematic representation of the map's distinctive features as inset. b) Normalized wavenumber $\tilde{k}_\mathrm{c}=k_\mathrm{c}d$; a zoomed-in view on the area where the upturn's sharpness drastically changes.}
\end{figure*}

Let us have a look at a first example of frequency and field map of caustic properties in Fig. \ref{fig_ex_fieldfreq_maps}. In the presented graphs, the red color means that either the corresponding $(h,\nu)$ point was not investigated because its reduced field is above the reduced FMR field $h_\mathrm{FMR}$, or because no caustic points were found there.

First of all, one can see that there is indeed a portion of the $(h, \nu)$ plane where no caustic points exist. This occurs for frequencies above a certain $\nu_\mathrm{m}(h,\eta)$. Then, going down in reduced frequency, there appears to be an oblique boundary  between two regions of the map. Above it, $\tilde{k}_\mathrm{c}$ quickly enters the dipole-exchange regime, which we will only present but not discuss quantitatively as it corresponds to a situation where our model is less reliable. Below the boundary, the reduced caustic wavenumber is much smaller than 1. Correspondingly, a boundary which we will label $\nu_\mathrm{b}(h,\eta)$ appears at the same position on the plot of $\varphi_\mathrm{c}$; this angle also seems close to constant over much of the region below the boundary. In both cases, its sharpness decreases towards low $h$, and at vanishing reduced field the transitions in $\tilde{k}_\mathrm{c}$ or $\varphi_\mathrm{c}$ are both smooth. All these features are represented on a simplified representation of the map of $\tilde{k}_\mathrm{c}$ shown as inset on the $\varphi_\mathrm{c}$ map, including the point $(h_\mathrm{c},\nu_\mathrm{c})$ at which the sharp boundary seems to end. A zoomed-in view on ($h_\mathrm{c}$,$\nu_\mathrm{c}$) is shown in the inset of Fig. \ref{fig_ex_fieldfreq_maps}.b).

In the following, we will refer to the lowest reduced field at which this boundary is sharp as $h_\mathrm{c}$ and denote $\nu_\mathrm{c}=\nu_\mathrm{b}(h_\mathrm{c},\eta)$. As we shall see in more details, this abrupt boundary corresponds to a change in the number of caustic points by two. The lowest point $(h_\mathrm{c},\nu_\mathrm{c})$ is actually a cusp in the domain of existence of the two additional caustic points. We point out that for all reduced fields and frequencies, the maps shown in Fig. \ref{fig_ex_fieldfreq_maps} displays the lowest caustic wavenumber respectively the associated wavefront angle.

\begin{figure*}[!h]
\includegraphics[width=14.59cm]{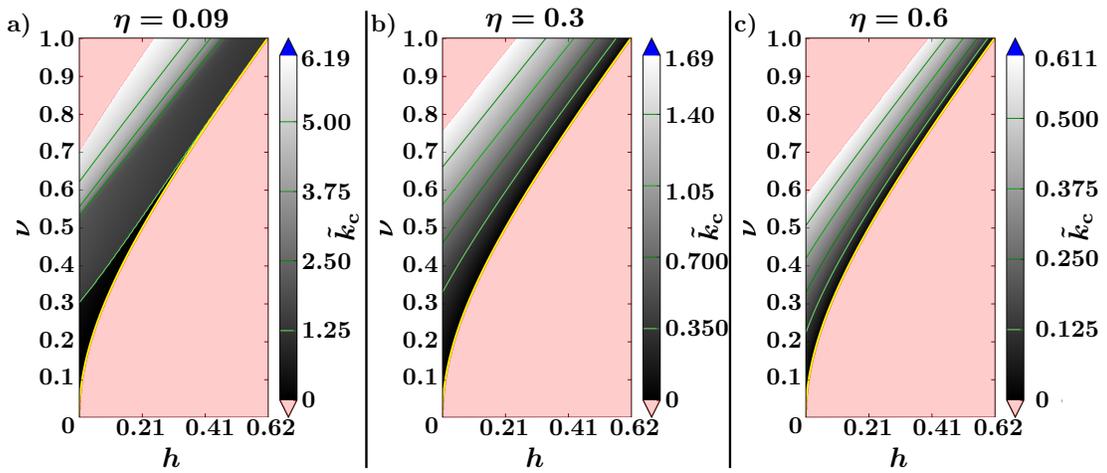}
\caption{\label{fig_otherEtaVals_examples}Examples of field and frequency maps for a) $\eta=0.09$, b) $\eta=0.3$, and c) $\eta=0.6$; only the reduced caustic wavenumber is shown.}
\end{figure*}

Before moving on to discussing the low-frequency pocket, its boundary and the existence of additional caustic points, and finally the threshold frequency for the absence of caustic points, we stress that the behaviour of caustics strongly depends on $\eta$. As an example, we show in Fig. \ref{fig_otherEtaVals_examples} field and frequency maps for $\eta=0.09, 0.3, 0.6$ (from left to right). At the lowest value, the boundary $\nu_\mathrm{b}$ extends all the way to $h=0$, whereas the two other maps do not display such a sharp behaviour. In addition to the expected changes in range of values for $\tilde{k}_\mathrm{c}$, one can see that the overall shape of the domain of existence of CSWBs also changes. From here on, we will call this area $\mathcal{D}$. From $\eta=0.09$ to $0.3$, we see that $\mathcal{D}$ has expanded in the vertical direction at low $h$. In even thinner films, for $\eta=0.6$, the average slope of $\nu_\mathrm{m}(h,\eta)$ has not changed much, yet $\nu_\mathrm{m}(0,\eta)$ has decreased; as a result, $\mathcal{D}$ shrinks vertically.

\begin{figure*}[!h] 
\includegraphics[width=14.59cm]{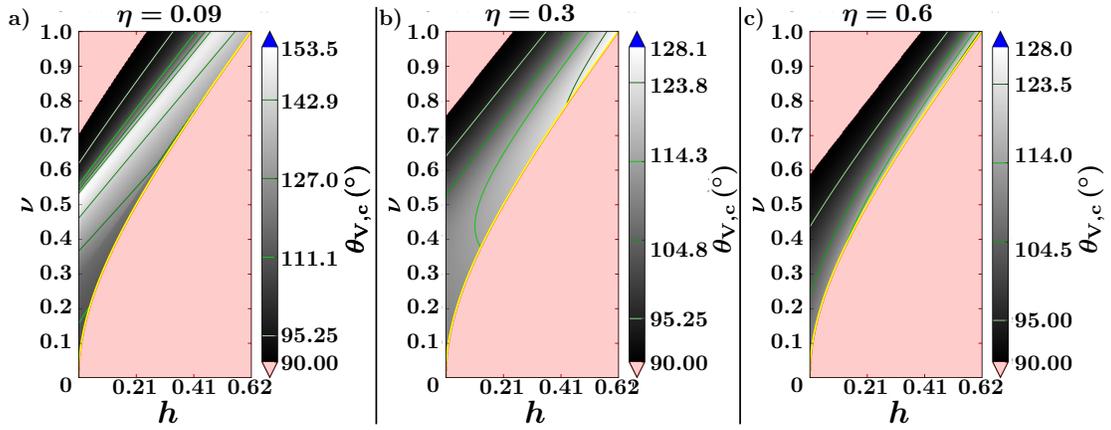}
\caption{\label{fig_otherEtaVals_examplesThetaV}Examples of field and frequency maps for the CSWB direction $\theta_\mathrm{V,c}$, at a) $\eta=0.09$, b) $\eta=0.3$, and c) $\eta=0.6$.}
\end{figure*}

By contrast, even if the caustic group velocity direction displays a similar wealth of features as the caustic wavefront angle and reduced wavenumber, the jumps across the boundary $\nu_\mathrm{b}$ are much less significant when they exist. An example of this is shown in Fig. \ref{fig_otherEtaVals_examplesThetaV}, which shows maps for $\theta_\mathrm{V,c}$ at the same values of $\eta$ as in Fig. \ref{fig_otherEtaVals_examples}.

In a certain range of reduced dipolar-exchange length, we find that there may actually be more than one caustic point on the slowness curve. Empirically, we observe that the additional caustic points may exist for $\tilde{k}_\mathrm{c}<1$. When this inequality holds, the number of caustic points is either equal to one or to three; two being possible but only on a 1D curve in the field and frequency plane; this curve includes the aforementioned boundary $\nu_\mathrm{b}$. Qualitatively, this is due to the fact that in the corresponding range of field and frequency, when $\mathrm{d}\theta_\mathrm{V}/\mathrm{d}\tilde{k}$ crosses 0, it does so with a local behaviour somewhat reminiscent of a polynomial of the type $P(\tilde{k};a,b)=(\tilde{k}-\tilde{k}_\mathrm{c})^3+a\cdot (\tilde{k}-\tilde{k}_\mathrm{c})+b$, where $a$ and $b$ are real parameters. If $a>0$, there exists only one root, whereas if $a<0$ and $|b|$ is sufficiently small, there exists three distinct roots. 

 The domain in the field and frequency plane with these three roots will be referred to as $\mathcal{D}_3$ from now on, by contrast with $\mathcal{D}_1=\mathcal{D}\setminus\mathcal{D}_3$ in which there is only one caustic point instead of three. We will now describe $\mathcal{D}_3$ using the $P(\tilde{k};a,b)$ approximant to $\mathrm{d}\theta_\mathrm{V}/\mathrm{d}\tilde{k}$ for the sake of simplicity.

\begin{figure*}[!h]
\includegraphics[width=14.59cm]{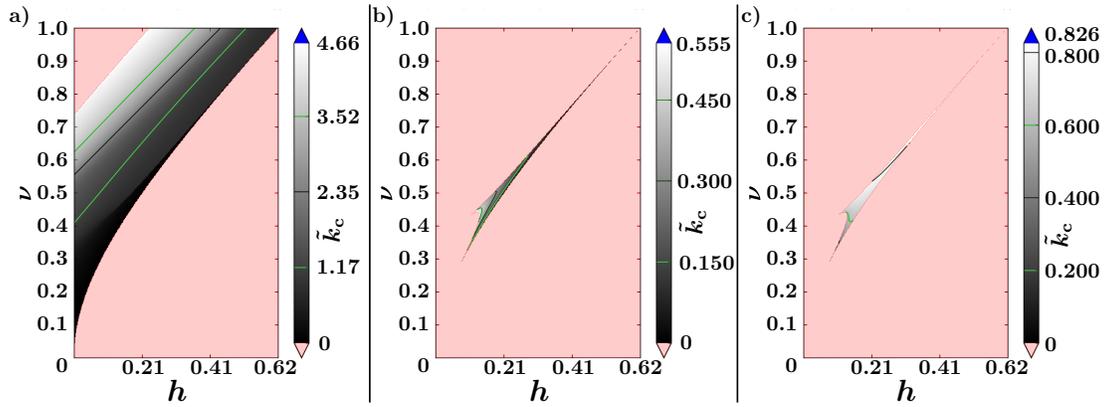}
\caption{\label{fig_3CPs}Field and frequency maps for $\eta=0.12$, looking at the three reduced caustic wavenumbers. Note the distinct grey scales for each graph. a) Lowest $\tilde{k}_\mathrm{c}$ in the presence of several caustic points, and single value for $\tilde{k}_\mathrm{c}$ otherwise. b) Intermediate value for $\tilde{k}_\mathrm{c}$ if several caustic points exist. c) Largest reduced caustic wavenumber.}
\end{figure*}

Let us start with Fig. \ref{fig_3CPs}, which displays the same field and frequency map for $\tilde{k}_\mathrm{c}$ as in Fig. \ref{fig_ex_fieldfreq_maps} along with the maps for the two other reduced caustic wavenumbers. The two additional solutions can be shown to coincide on the rounded boundary of $\mathcal{D}_3$ to the lower left, which will be referred to as $\partial \mathcal{D}_{3,l}$. Entering $\mathcal{D}_3$ through this boundary by increasing $\nu$ corresponds to the situation where $|b|$ becomes small enough to allow the two additional caustic points (with respect to the one with lowest $\tilde{k}_\mathrm{c}$), thanks to $a$ being negative enough. Increasing $h$ on the other hand mostly decreases $a$: upon crossing  $\partial \mathcal{D}_{3,l}$, a pair of caustic points with higher $\tilde{k}_c$'s appears. Of course, exactly on $\partial \mathcal{D}_{3,l}$ the two additional roots of $\mathrm{d}\theta/\mathrm{d}\tilde{k}$ are identical.

Starting from inside $\mathcal{D}_3$, if one increases the reduced frequency, eventually the caustic point with the intermediate value of $\tilde{k}_\mathrm{c}$ merges with the one featuring the smallest reduced wavenumber. This happens on the other boundary of $\mathcal{D}_3$, which we will call $\partial\mathcal{D}_{3,u}$ from now on. This situation corresponds to $\nu=\nu_\mathrm{b}(h,\eta)$. Just above this boundary, the value of $b$ is low enough so that  only one root of $\mathrm{d}\theta/\mathrm{d}\tilde{k}$ remains. That is the reason for the discontinuity in $\tilde{k}_\mathrm{c}$ in Fig. \ref{fig_ex_fieldfreq_maps}: the lowest caustic wavenumber jumps to what was the highest of the three $\tilde{k}_\mathrm{c}$'s below $\nu_\mathrm{b}$. Experimentally, this could imply that SW excitation around this threshold wavenumber would have marked changes in intensity as a function of frequency.

 Based on the above, since the two boundaries other than ferromagnetic resonance each imply that a \emph{different} pair of caustic points coincide, we can infer that on the cusped intersection of $\partial\mathcal{D}_{3,l}$ and $\partial\mathcal{D}_{3,u}$, there exists a single caustic point corresponding to three of them coinciding on the slowness curve. This is precisely the point $(h_\mathrm{c}, \nu_\mathrm{c})$ from the inset in Fig. \ref{fig_ex_fieldfreq_maps}.

It is important to note that while a purely mathematical analysis yields well-defined, separate caustic points, experimentally the distinction between close caustic points may well be impossible. In fact, there exists no straightforward experimental signature of $\mathrm{d}\theta_\mathrm{V}/\mathrm{d}\tilde{k}$ crossing 0, and portions of the slowness curve where this derivative is small but non-zero can behave similarly to an actual caustic point, as was noted by Gallardo \emph{et al.} \cite{Gallardo2021_SWcausticsInSynthAntiferromagnet}. Nevertheless, the presence of more than one caustic point constrains a slowness curve to be almost straight in their vicinities; this should then favour marked caustics.

\subsection{Low-frequency pocket}\label{sec_LowFreqPocket}

The low-frequency regime is important as it corresponds to a well established domain of validity of our theoretical model as well as wavelengths which can still be excited and detected reasonably easily in experiments.

\subsubsection{Analytics}

As could be seen in Fig.\ref{fig_otherEtaVals_examples}, the shape or even the existence of the low-frequency pocket strongly depends on the chosen value of $\eta$. Nevertheless, we can investigate the behaviour of caustics there by taking the limit $\nu\rightarrow 0$. In order to remain below ferromagnetic resonance, we also take the limit $h\rightarrow0$. Assuming $h=0$ simplifies the computation of the quantity $\tan{\theta_\mathrm{V}}=\tan{\varphi}\cdot[1+f(\tilde{k},\nu,\eta)]$, where $f$ is a function given in the Supplementary Materials. We can then differentiate this with respect to $\tilde{k}$, take the limit $\nu\rightarrow 0$ and Taylor-expand the derivative; the details are provided in the Supplementary Materials. Eventually, we find that:

\begin{equation}
\tilde{k}_\mathrm{c}(\nu\rightarrow0)=3\nu^2+\mathcal{O}(\nu^4).
\end{equation}

It was expected that the caustic wavenumber goes to zero; we can furthermore show that the lowest reduced wavenumber on the slowness curve (still in zero applied field) \emph{i.e.} the Damon-Eshbach wavenumber goes to zero as:

\begin{equation}
\tilde{k}_\mathrm{min}(\nu\rightarrow0)=2\nu^2+\mathcal{O}(\nu^4)
\end{equation} \\ which proves that CSWBs exist down to vanishing reduced frequencies, regardless of their values. In this limit, the associated caustic wavefront angle is such that:

\begin{equation}
\cos{\varphi_\mathrm{c}}=\frac{1}{\sqrt{3}}+\mathcal{O}(\nu^2).
\end{equation}

From the latter, we also get the CSWB direction $\theta_\mathrm{V,c}$:

\begin{equation}
\tan\theta_V(\tilde{k}_\mathrm{c},h\rightarrow 0,\nu\rightarrow0)=-2\sqrt{2}+\mathcal{O}(\nu^2)
\end{equation}

The strength of this result lies with its independence on $\eta$; this is not surprising as in the limit we are considering, the CSWB's wavelength diverges which means it must be much larger than both the film thickness $d$ and the dipolar-exchange length $l_\mathrm{ex}$, however large they may be. The numerical values for the limits of $\varphi_\mathrm{c}$ and $\theta_\mathrm{V,c}$ are \emph{ca.} \ang{54.74} and \ang{109.5}, respectively.

\subsubsection{Comparison with literature}

We present in Table~\ref{tab_CompPredLit} a comparison between experimental reports on caustics and predictions we make for the same conditions, focusing on the CSWB direction. Whenever there are three caustic points, the indicated predicted value for $\theta_{\mathrm{V},\mathrm{c}}$ is the closest found across all three caustic points.

%\begingroup
%\squeezetable
\begin{table*}
\caption{\label{tab_CompPredLit}Comparison between reports on CSWBs and our predictions for the beam direction $\theta_{\mathrm{V,c}}$.}
\begin{ruledtabular}
\begin{longtable}{cp{3.5cm}cccccc}
Ref. & Excitation method & Material (thickness in nm) & Predicted $\theta_{\mathrm{V},\mathrm{c}}$ & Measured $\theta_{\mathrm{V},\mathrm{c}}$ & $h$ & $\nu$ & $\eta$  \\ 
\hline 
\cite{Sebastian2013_CausticSWbeamsFromEdgeModeInHeuslerAlloy} & Edge modes of a waveguide and nonlinearities & Co$_2$Mn$_{0.6}$Fe$_{0.4}$Si (30)  &  \ang{113}  &  \ang{123} & 3.81$\cdot10^{-2}$  & 0.287  & 0.15 \\ %nu=0.2873
\hline 
\cite{Bertelli2021_NVcenterImagingOfSWsAndCaustics} &   Corners of slotline termination and scattering off a defect &  YIG (235) & \ang{123} &  \ang{124}, \ang{122} &    0.126 &  0.427 & 7.36$\cdot10^{-2}$ \\ %thetaV predicted\ang{123.3071}, measured ones \ang{124.3} \ang{121.5}, h=0.126408, nu=0.42715871, eta=0.073617
\hline
\cite{Bertelli2020_MultipleSWcausticsFromStriplineCorners} & Corners of slotline termination & YIG (245) & \ang{119} & \ang{118} &  0.126 & 0.427 & 7.06$\cdot10^{-2}$ \\%eta=0.07062, nu=0.4272, h=0.12613, prediced theta_V \ang{118.83},measured one \ang{117.93} 
\hline
\cite{Gieniusz2013_AntidotForDiffractingSWs_BLS} & Spin wave scattering off antidots & YIG (4.5$\cdot$10\textsuperscript3) & \ang{169} & \ang{128} & 0.557 & 0.939 & 3.84$\cdot10^{-3}$ \\ %predicted theta_V \ang{169.4}, measured \ang{128.3}
\hline
\cite{Kostylev2011_CausticSWbeamsFromDecayingSWwavepacket} & Collapsing spin-wave bullet &  YIG (5$\cdot$10\textsuperscript3) & \ang{137} & \ang{137}  & 1.040 & 1.442 & 3.46$\cdot10^{-3}$ \\%nu=1.4423, h=1.040, predicted theta_V \ang{137.4364}, measured \ang{136.5}
\hline
\cite{Buettner2000_LinAndNonLinSWscatteringInYIGbyBLS} & Spin wave scattering off a defect &  YIG (7$\cdot$10\textsuperscript3) & \ang{139} &   \ang{135} &  2.47$\cdot10^{-3}$  & 1.616 & 2.47$\cdot10^{-3}$  \\ %nu=1.61551, h=0.002471, predicted theta_V \ang{138.655}, measured \ang{134.521}
\hline
\end{longtable}
\end{ruledtabular}
\end{table*}
%\endgroup

We find a reasonable agreement in quite a few cases, generally for the larger values of $\eta$ (\emph{i.e.} for thinner films) with the notable exception of the report by Sebastian \emph{et al.} \cite{Sebastian2013_CausticSWbeamsFromEdgeModeInHeuslerAlloy}. However, in this case, the theoretical dispersion relation that we use may not be accurate any more due to the strong lateral confinement of spin waves.

\begin{figure}[h!]
\includegraphics[width=8.599cm]{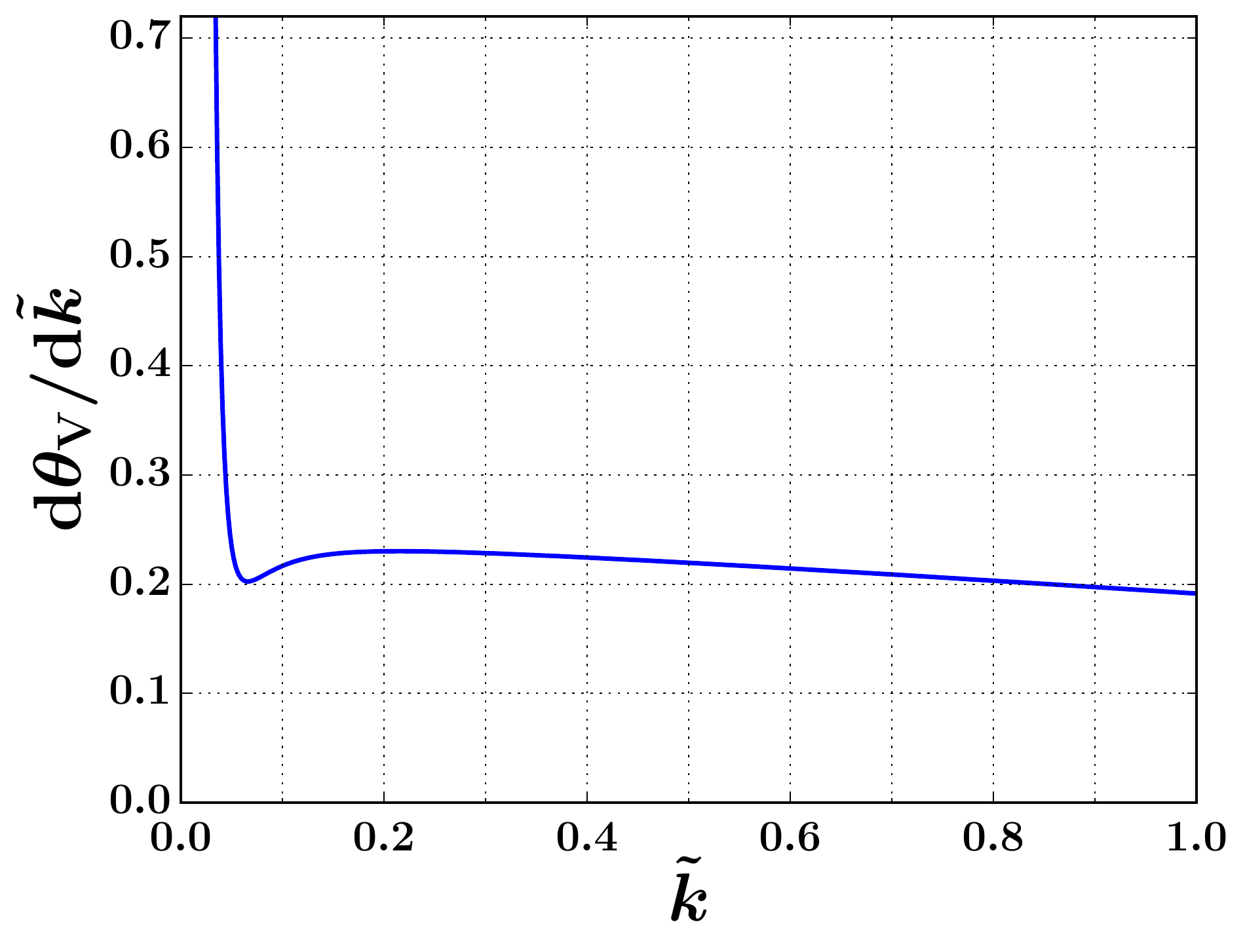}
\caption{\label{Fig_Gieniusz2013_dthetadx}Calculated derivative of the group velocity direction with respect to the reduced wavenumber in the \SI{4.62}{\giga\hertz} spin wave excitation described by Gieniusz \emph{et al.} \cite{Gieniusz2013_AntidotForDiffractingSWs_BLS}.}
\end{figure}

Furthermore, we find much larger discrepancies in several cases. For instance, if we consider the excitation of a caustic-like beam by Gieniusz \emph{et al.} \cite{Gieniusz2013_AntidotForDiffractingSWs_BLS} at \SI{4.62}{\giga\hertz} and under an induction of \SI{98}{\milli\tesla} in a \SI{4.5}{\micro\meter} thick YIG film, our model predicts a caustic point at reduced wavenumber $13.2$, with a beam direction \ang{169}. However, the relevant reduced wavenumbers in this experiment are in the range of a few percents \cite{Gieniusz2013_AntidotForDiffractingSWs_BLS}, and the measured beam direction is \ang{128}. The origin of this strong disagreement is easily understood by observing the derivative $\mathrm{d}\theta_\mathrm{V}/\mathrm{d}\tilde{k}$ in this case. As Fig. \ref{Fig_Gieniusz2013_dthetadx} reveals, there exists a local minimum at $\tilde{k}\simeq 0.0659$ for $\mathrm{d}\theta_\mathrm{V}/\mathrm{d}\tilde{k}$ deep in the dipolar-dominated regime. Moreover, the associated group velocity direction is \ang{128}, and past the next local maximum, similar values of $\mathrm{d}\theta_\mathrm{V}/\mathrm{d}\tilde{k}$ are reached again only for $\tilde{k}\simeq 0.9$. This illustrates the impossibility to distinguish a close-to-straight slowness curve from a true caustic point from measurements alone. %Gieniusz2013_AntidotForDiffractingSWs_BLS predicted wavenumber 13.18292

Discrepancies may also arise due to the source's non-ideal excitation efficiency, for instance if it is too directional. This is illustrated by the excitation of caustic-like spin wave beams by K\"orner \emph{et al.} \cite{Koerner2017_DEcausticSWbeamsFromNotchedCPWs}. One of the reported TR-MOKE measurements deals with a \SI{60}{\nano\meter} thin permalloy film driven at an excitation frequency of \SI{16.08}{\giga\hertz}, under \SI{160}{\milli\tesla} applied induction; the authors observe twin beams with a wavefront angle of \ang{65}, a beam direction \ang{114}, and a reduced wavenumber of 0.314. Yet, the expected caustic spin wave beams in these conditions should feature a reduced wavenumber of 1.7063, a beam direction \ang{138.62}, not to mention a wavefront angle of \ang{53.27}. In this case, it appears that the excited spin waves simply correspond to the rather narrow portion of the slowness curve that could be excited by the authors' tapered coplanar waveguide segments \footnote{We point out that the segment perpendicular to the tapered waveguide segments was at an angle of about \ang{60} with respect to the applied field in these experiments \cite{Koerner_PhD_Thesis}}. Indeed, at the measured wavefront angle of \ang{65}, in the authors' experimental conditions, the expected reduced wavenumber is about 0.28 (which falls rather far from zeroes in the antenna's expected excitation efficiency \cite{Koerner_PhD_Thesis}), and the beam direction \ang{120.2}. We do not have an explanation for the remaining deviation in beam direction, though.

\subsubsection{Experimental results}

We now present results from experiments we have carried out in order to validate our theoretical approach. Our aim here is to measure CSWBs and compare their properties with our predictions. In order to access CSWBs experimentally, the reciprocal-space Fourier components of its magnetic field must span a broad range of wave vectors. The ideal situation where all wave vectors are accessible corresponds to an unrealistic point source, which can obviously not correspond to any high-frequency antenna. As a result, we choose a compromise between ease of fabrication, and broad-band excitation efficiency, namely a half-ring shaped stripline antenna. This design allows for a spin wave excitation of the slowness curve within $\varphi \in [0,\pi]$, \emph{i.e.} twice the quadrant previously investigated. Of course, this excitation is not uniform because of the microwave antenna dimensions on the order of a micrometer.  

\begin{figure}
	\includegraphics[width=4.5cm]{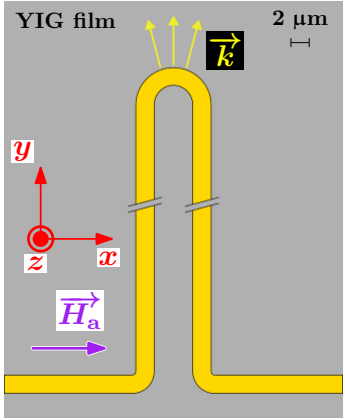}	
	\caption{\label{fig:measurement_geometry}Schematic of the measurement geometry. The half-ring shaped antenna excites spin wave propagation within a broad angular spectrum.}
\end{figure}

Our experiments were carried out using Time-Resolved Magneto-Optical Kerr Effect (TR-MOKE) microscopy. Here, the dynamic out-of plane component of the magnetization $\delta m_z$ is spatially mapped in the $xy$-plane at a fixed phase between the microwave excitation frequency and the laser probing pulses. This enables direct imaging of the spin wave propagation in the magnetic film. The wavenumber resolution of the set-up lies within the dipolar-dominated regime. Indeed, our spatial resolution $r$ is about \SI{0.29}{\micro\meter} (see Supplementary Materials), so that for a film thickness $t\sim$\SI{100}{\nano\meter}, the largest accessible reduced wavenumbers are $2\pi/(2r)\cdot t\sim 1$.

It shall be noted that the position of the microwave antenna in the resulting Kerr images is extracted from the topography image which is acquired simultaneously and is proportional to the reflectivity of the sample. Further information on TR-MOKE can be found in the Supplementary Materials. These experiments were performed on a \SI{200}{\nano\meter} thick YIG film grown on a gadolinium gallium garnet (GGG) substrate using liquid phase epitaxy. Considering this materials' parameters \cite{Klingler2014_fmrMeasurementOfYIGexchangeCst}, if not stated otherwise, $\eta=0.087$ for all measurements. On top of the YIG film the \SIrange{2}{3}{\micro\meter} wide microwave antenna was patterned by optical lithography with subsequent Ar-presputtering and electron-beam-induced evaporation of Cr(\SI{5}{\nano\meter})/Au(\SIrange{100}{220}{\nano\meter}). During the measurement the external bias field $\overrightarrow{H_\text{a}}$ was always kept fixed such that it aligned with the legs of the antenna structure along the $x$-direction. A sketch of the measurement geometry can be found in Fig.\,\ref{fig:measurement_geometry}. At this stage, we point out one complication resulting from this design. When driving the antenna with a microwave field, the legs themselves excite spin waves in the Damon-Eshbach geometry \cite{Damon_Eshbach1961_SWmodesInThinFilms__RefPaper}. These modes are not of interest for the generation of CSWBs, but due to the relatively long attenuation length in YIG \cite{Bertelli2020_MultipleSWcausticsFromStriplineCorners} they may propagate to the tip of the antenna and interfere with the spin waves excited by the half-ring. In order to suppress this effect, two different approaches where applied. Either the length of the antenna was set to \SI{50}{\micro\meter} and the YIG between the legs and tip was etched away, or the antenna was patterned to be \SI{1}{\milli\meter} long in the first place.

\begin{figure}[!h]
	\includegraphics[width=6.5cm]{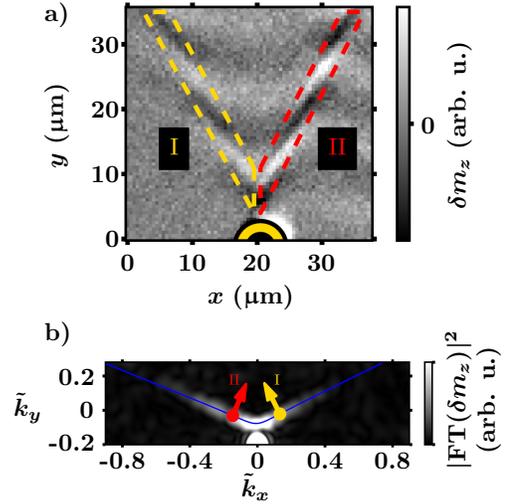}
	\caption{\label{fig:1.44GHz_5mT}Measurement data obtained for $\eta = 0.087$, $h=0.028$ and $\nu=0.292$. a) Kerr image acquired from TR-MOKE. Two spin wave beams highlighted in yellow and red propagate from the tip of the antenna. b) Squared modulus  of the Fourier transform (FT) of the Kerr image and expected slowness curve (blue). The yellow and red points and arrows indicate the expected caustic points and their respective group velocity directions. Caustic points I and II correspond to beams I and II in the Kerr image.}
\end{figure}

The first Kerr image shown in Fig.~\ref{fig:1.44GHz_5mT}.a) was obtained at a constant microwave frequency $f=$\SI{1.44}{\giga\hertz} and an external field $\mu_0H_\text{a}=$\SI{5}{\milli\tesla}. This corresponds to $h$~=~0.028, $\nu$~=~0.292. The width of the waveguide was \SI{2}{\micro\meter} and the distance between the legs and the tip was \SI{1}{\milli\meter}. In the spatial map, two spin wave beams with well-defined propagation directions are visible; moreover, the phase and group velocities are clearly non-collinear to each other. Here, beam II stems from the waveguide excitation in the quadrant $\varphi \in [\pi/2,\pi]$. The beam angles of beams I and II with respect to the positive $x$ direction are found to be $119.00^\circ$ (beam I) and $64.28^\circ$ (beam II) which results in effective beam directions of $\theta_\text{I}=119.00^\circ$ and $\theta_\text{II}=180^\circ-64.28^\circ=115.72^\circ$, respectively. The discrepancy between $\theta_\text{I}$ and $\theta_\text{II}$ simply originates from a small misalignment of the external field with respect to the waveguide legs. Since $\overrightarrow{H_\text{a}}$ is not fully parallel to the $x$-axis, the slowness curve is rotated by a small angle $\alpha_\text{H}=(\theta_\text{I}-\theta_\text{II})/2 \approx 1.64^\circ$ in our frame of reference. Keeping this in mind, we extract an average beam direction $\theta_\text{V,e}=117.36^\circ$, a wavefront angle $\varphi_\text{e}=50.66^\circ$ and a reduced wavenumber $\tilde{k}_\text{e}=0.211$. These experimental findings are in good agreement with our theoretical approach; indeed, values of $\theta_\text{V,c}=115.05^\circ$, $\varphi_\text{c}=51.29^\circ$ and $\tilde{k}_\text{c}=0.223$ are predicted for a CSWB in our experimental conditions.

We can obtain further insight in reciprocal space with the Fourier-transformed (FT) data shown in Fig.~\ref{fig:1.44GHz_5mT}.b). Generally speaking, the FT data allows for a direct observation of the slowness curve in $\tilde{k}$-space. In order to reduce spectral leakage, a Hanning windowing was applied; the latter provides a good trade-off between frequency and amplitude accuracy. We see that the chosen antenna structure indeed excites a wide range of wave vector directions. The gaps in the spectrum arise from the finite antenna dimensions, as previously mentioned. We find a good agreement between the slowness curve (blue curve) derived from our model (and corrected by the external field angle $\alpha_\text{H}$). More importantly, this graph confirms that the antenna structure grants access to the expected caustic points (yellow and red points) since the Fourier magnitude is still sufficiently large in that region. To conclude, caustic points I and II can be assigned to beams I and II from the Kerr image.

We may now turn to the additional caustic points predicted by our model. The chosen triplet $(\eta, h, \nu)$ is an element of the $\mathcal{D}_3$ set, and we would expect two further caustic points $\theta_\text{V,c,2}$~=~$113.74^\circ$, $\varphi_\text{c,2}$~=~$33.00^\circ$, $\tilde{k}_\text{c,2}$~=~$0.662$ and $\theta_\text{V,c,3}$~=~$114.02^\circ$, $\varphi_\text{c,3}$~=~$28.78^\circ$, $\tilde{k}_\text{c,3}$~=~$1.227$. These reduced wave vectors could actually be resolved by our experimental set-up where $\tilde{k}_\text{res}$~$\approx$~$2.2$. The reciprocal space image in Fig.~\ref{fig:1.44GHz_5mT}.b), however, displays a very low amplitude for $\tilde{k}\gtrsim0.55$ meaning that the microwave antenna cannot excite the other caustic points very efficiently. Hence, only the low frequency pocket can be accessed.

\begin{figure}[!h]
	\includegraphics[width=8.599cm]{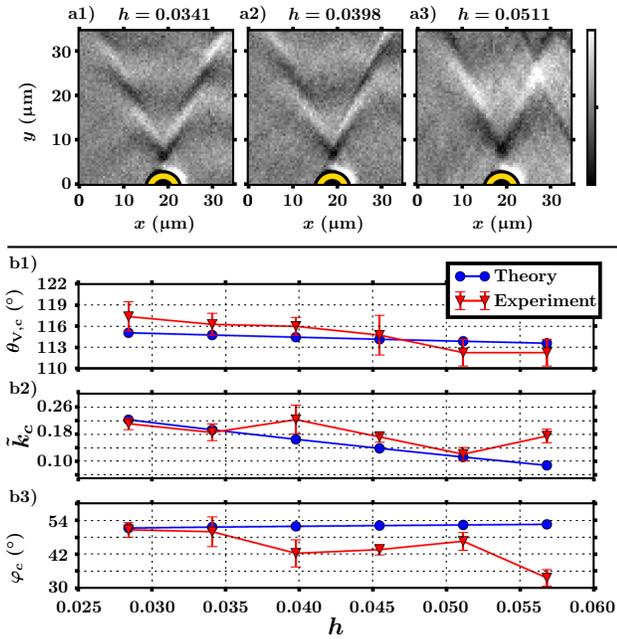}
	\caption{\label{fig:complete_analysis}Measurement data obtained for $\eta = 0.087$ and $\nu=0.292$. Upper part: acquired Kerr images for reduced fields of a1) $h=0.0341$, a2) $h=0.0398$, and a3) $h=0.0511$,. b1-3) comparison between experiment and theoretical predictions of caustic point properties $\theta_\text{V,c}$, $\tilde{k}_\text{c}$, $\varphi_\text{c}$. The error bars are the standard deviations from a bootstrapping fit procedure.}
\end{figure}

Further Kerr images were taken for the same $\nu$, but for different $h$ values. The $h$ values were chosen such that they lie beneath the expected FMR field $h_\text{FMR}\approx0.078778$. A selection of the resulting Kerr images is illustrated in the upper part of Fig.\,\ref{fig:complete_analysis}. In each of them, twin spin wave beams are apparent. An overview of all the beam properties for the corresponding $h$ values is plotted in the lower part of Fig.\,\ref{fig:complete_analysis}. Here, the relevant parameters from every individual beam are extracted with image processing and bootstrapping least squares regression procedures. An example on how one set of experimental data points is obtained can be found in the Supplementary Materials. The reasonable, sometimes even very good agreement between predicted and experimental values of $\theta_\text{V,c}$ and $\tilde{k}_\text{c}$ strongly suggests true CSWBs. The deviation of the beam directions is mostly within the range of the external field angle. The larger discrepancy between predicted and measured wavefront angles $\varphi_\text{c}$ is attributed to the narrowness of the CSWB.

\begin{figure}[!h]
	\includegraphics[width=8.59cm]{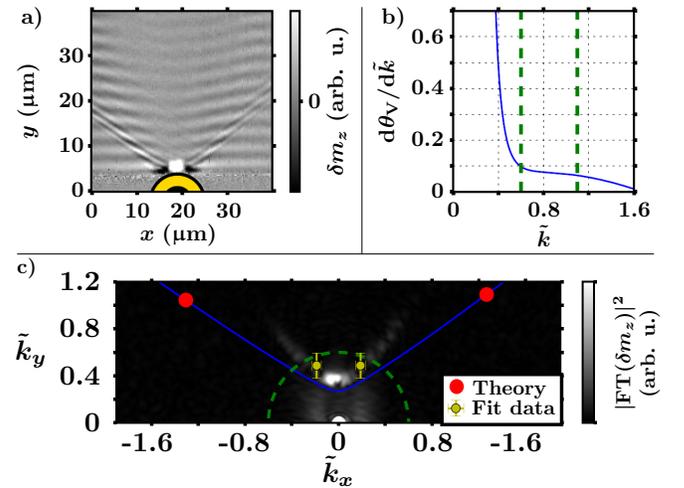}
	\caption{\label{fig:1.84GHz_5mT}a) Kerr image with twin beams obtained with $\eta = 0.087$ $h=0.028$ and $\nu=0.372$. b) Calculated derivative of the group velocity direction with respect to the reduced wavenumber. Dashed green lines highlight close-to-straight slowness curve. c) FT of Kerr image. The experimentally observed beam parameters are depicted in yellow, the  calculated slowness curve in blue and the calculated caustic points in red. Dashed green semicircle illustrates lower limit of close-to-straight portion of slowness curve.}
\end{figure}

Beam-like features which do not coincide with a caustic point were detected as well. This time, the measurements were conducted with the \SI{50}{\micro\meter} antennna structure and partially etched film. The width of the antenna was \SI{3}{\micro\meter}. The resulting Kerr map for $f=$\SI{1.84}{\giga\hertz} ($\nu=0.372$) and $\mu_0H_\text{a}=$\SI{5}{\milli\tesla} ($h=0.028$) is shown in the left upper half of Fig.\,\ref{fig:1.84GHz_5mT}. In this geometry, a Damon Eshbach-like mode propagating from the YIG edge could not be fully suppressed; it is visible as a plane wave background. Our procedure to analyze spin wave beams yields $\theta_\text{V,e}=136.33^\circ$, $\varphi_\text{e}=68.97^\circ$ and $\tilde{k}_\text{e}=0.522$, whereas our model predicts a caustic point with $\theta_\text{V,c}=121.39^\circ$, $\varphi_\text{c}=35.84^\circ$ and $\tilde{k}_\text{c}=1.564$. 

The origin of the experimentally observed beams may be twofold. Firstly, a close-to-straight slowness curve similar to the case of Gieniusz \textit{et al.} \cite{Gieniusz2013_AntidotForDiffractingSWs_BLS} is predicted to exist within relatively close distance to $\tilde{k}_\text{e}$. The $\mathrm{d}\theta_\text{V}/\mathrm{d}\tilde{k}$ plot in Fig.\,\ref{fig:1.84GHz_5mT}.b) displays almost a constant behaviour between $0.6$~$\lesssim$~$\tilde{k}$~$\lesssim$~$1.2$ (marked with green dashed lines). The proximity of the experimental caustic point to a straight-to-close slowness curve is also illustrated in the FT data in the lower part of Fig.\,\ref{fig:1.84GHz_5mT}. Here, the dashed green semicircle represents the lower bound of $\tilde{k}=0.6$ and the extracted beam points are highlighted in yellow. For this portion of the slowness curve, group velocity directions of up to $121.39^\circ$ are predicted. This beam direction, however, is still in stark contrast with the measurement result. Moreover, the calculated slowness curve (blue curve) deviates significantly from the FT data. The difference between reciprocal space image and our model may show the limit of the model applicability, since a film with $\eta = 0.087$ may not be considered a thin film anymore. This results in predictions which are less reliable at higher $\nu$ values. A second possible origin of the beams is the excitation efficiency of the microwave antenna as there are many gaps in the FFT spectrum. The beams appear to be located close to some of them, and hence, may correspond to the excitation of only a small portion of the slowness curve within this region.

\subsection{Caustic point of higher order}

Based on the conclusions from section \ref{sec_general_features}, we know that the intersection of $\partial\mathcal{D}_{3,l}$ and $\partial\mathcal{D}_{3,u}$ there exists a single caustic point on the slowness curve; in the schematic discussion from the above based on the approximant $\mathcal{P}(\tilde{k};a,b)$, it corresponds to $a=0$ and $b=0$, which means that $\mathrm{d}\theta_\mathrm{V}/\mathrm{d}\tilde{k}\sim (\tilde{k}-\tilde{k}_\mathrm{c})^3$ around this point. To put it differently: at this intersection, corresponding to the cusp seen in Fig. \ref{fig_3CPs}, the caustic point is not a simple extremum for $\theta_\mathrm{V}$ on the slowness curve but an undulation point, in the vicinity of which $\theta_\mathrm{V}-\theta_{\mathrm{V},\mathrm{c}}\sim (\tilde{k}-\tilde{k}_\mathrm{c})^4$.

\begin{figure}%[!h]
\includegraphics[width=7.5cm]{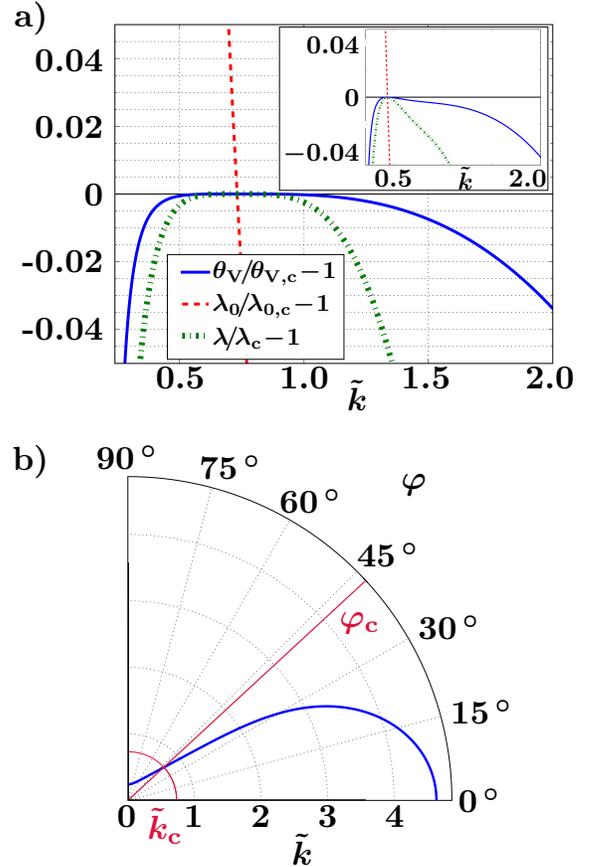}
\caption{\label{fig_relThetaV_LamNat_LamApp}a) Plots of the relative deviations from the following caustic point properties as a function of $\tilde{k}$: its group velocity direction $\theta_{\protect\mathrm{V}}$, its natural wavelength $\lambda_0=2\pi/\tilde{k}$ and its apparent wavelength $\lambda=2\pi/[\tilde{k}\cos{(\theta_{\protect\mathrm{V}}-\varphi)}]$. Main graph: $h=h_1=1.15\cdot10^{-21}$, $\nu=\nu_1=0.315279504$, $\eta=\eta_1=0.10253664614147$, which are extremely close to the values of $\nu_\mathrm{c}$ and $\eta$ for which $h_\mathrm{c}=0$. Inset: same $h$ and $\eta=\eta_1$, $\nu=0.95\cdot\nu_1=0.2995155288$. b) Slowness curve for $\nu_1$, $\eta_1$ and $h_1$; $\tilde{k}_\mathrm{c}\simeq 0.7259$. The slowness curve at $\nu_2$ is not shown for clarity, as it is very similar to the other one.}
\end{figure}

The existence of such an undulation point is of particular interest since the higher order in the dependence of  $\theta_\mathrm{V}$ on $\tilde{k}$ implies a flatter extremum in group velocity direction and therefore the possibility of larger portions of the slowness curve contributing to the CSWB. Moreover, as was discussed in Sec. \ref{sec_model}.\ref{sec_distinctiveFeats}, this does not necessarily mean an increase in spectral breadth of the CSWB since the latter depends on the apparent wavelength. In order to evidence this, we show in Fig. \ref{fig_relThetaV_LamNat_LamApp} how the group velocity direction as well as the natural and apparent wavelengths vary around a caustic point very close to one of higher order, here the one such that its corresponding critical field $h_\mathrm{c}$ is zero. The considered slowness curve corresponds to $h=h_1=1.15\cdot10^{-21}$, $\nu=\nu_1=0.315279504$, $\eta=\eta_1=0.10253664614147$.

Let us briefly outline how the coordinates $\nu_{\mathrm{c},0}=\nu_\mathrm{c}(h_\mathrm{c}=0)$ and $\eta_{\mathrm{c},0}=\eta_\mathrm{c}(h_\mathrm{c}=0)$ were found with a good accuracy. More details can be found in the Supplementary Materials. The starting point was a rough, hand-performed search for a value of $\eta$ bringing the cusp of $\mathcal{D}_3$ to lie on the ordinate axis in a field and frequency map. This yielded a starting point of $\eta_{\mathrm{c},0}^{(0)}=0.10$ and $\nu_{\mathrm{c},0}^{(0)}=0.31$. In these conditions, a caustic point was found for $\tilde{k}_{\mathrm{c},0}^{(0)}\simeq0.73$. We then began an iterative procedure using appropriate Taylor expansions of the dispersion relation and of an exact expression for $\theta_\mathrm{V}(h=0,\eta,\nu,\tilde{k},\varphi)$. Updating these at each step with the new solutions found by looking for the undulation point allows to converge to numerical values which we assimilate to the intersection of $\partial\mathcal{D}_{3,l}$ and $\partial\mathcal{D}_{3,u}$.

Over three iterations, the relative changes in the estimates steadily decrease in absolute value, from at most 5\% in the first step to at most $5\cdot10^{-6}$ in the last one, which provides the following guesses : $\tilde{k}_{\mathrm{c},0}^{(g)}=0.731717$, $\eta_{\mathrm{c},0}^{(g)}=0.1025366$, $\nu_{\mathrm{c},0}^{(g)}=0.3152796$. The latter can be compared with \emph{e.g.} the hand-refined values used for Fig. \ref{fig_relThetaV_LamNat_LamApp}: $\nu=\nu_1=0.315279504$, $\eta=\eta_1=0.10253664614147$, corresponding to $\tilde{k}_\mathrm{c}=0.725904$. It must be noted that the somewhat larger relative difference in terms of $\tilde{k}_{\mathrm{c},0}$ is due to the very steep dependence of  $\tilde{k}_\mathrm{c}(\nu,\eta,h\rightarrow0)$ on $\eta$. We do emphasize that the \emph{exact} location ($\nu_{\mathrm{c},0}$,$\eta_{\mathrm{c},0}$) is necessarily  different from ($\nu_1$, $\eta_1$) but close enough to highlight the qualitatively different behaviour of several characteristics of the slowness curve. Finally, we note that for the parameters from Fig. \ref{fig_relThetaV_LamNat_LamApp}, $\theta_\mathrm{V,c}=$\ang{118.36}, $\varphi_\mathrm{c}\simeq$\ang{42.75}, $\lambda_{0,\mathrm{c}}=84.41 l_\mathrm{ex}=8.655 d$, and $\lambda_{\mathrm{c}}\simeq 339.7 l_\mathrm{ex}=34.83 d$.

We now examine the properties of the caustic point of higher order in more detail. From Fig. \ref{fig_relThetaV_LamNat_LamApp}, the dependence of $\theta_\mathrm{V,c}$ and the apparent wavelength $\lambda$ on $\tilde{k}$ (in blue and green, respectively) clearly appears to be quartic rather than quadratic around the caustic point, which is where the deviations in natural wavelength (in red) go through 0. Its much steeper behaviour is easily understood by looking at the corresponding slowness curve in Fig. \ref{fig_relThetaV_LamNat_LamApp}.b): around $\tilde{k}_\mathrm{c}$ it is not only almost straight but the angle $\gamma$ between $\overrightarrow{\tilde{k}}$ and $\mathrm{d}\overrightarrow{\tilde{k}}/\mathrm{d}s$ is low, $\gamma\simeq$\ang{14.38}. Hence, since $\mathrm{d}(\tilde{k}^2)/\mathrm{d}s$ is large, $\lambda_0\propto 1/\tilde{k}$ varies fast.

By contrast, one can show that in the Taylor expansion of $\lambda$ in $(s-s_\mathrm{c})/\tilde{k}_\mathrm{c}$ around $\lambda_\mathrm{c}$, the first coefficient is always exactly zero at a caustic point. We stress again that this is caused by an unchanging projection of $\overrightarrow{k}$ on $\overrightarrow{e_\mathrm{g}}$ across the caustic point. If it is of higher order, it may be shown (see Supplementary Materials) that in this term, the contributions due to the second- and third-order variations of $\varphi$ and to those of $\tilde{k}$ cancel out. To put it differently, the projection $k\cdot\cos{(\theta_\mathrm{V}-\varphi)}$ is now constant up to fourth order in $(s-s_\mathrm{c})/\tilde{k}_\mathrm{c}$. On the other hand, if the considered caustic point is a regular extremum for $\theta_\mathrm{V}$, the term $\propto (s-s_\mathrm{c})^2$ will be non-zero.

To summarize the above paragraph: for geometrical reasons, the caustic point of higher order suppresses the quadratic and cubic variations of the apparent wavelength around $\lambda_\mathrm{c}$. Hence, $\lambda$ has then a markedly quartic behaviour at a caustic point of higher order. Furthermore, we point out that even a small offset in frequency makes it display a clearly quadratic behaviour. This is shown in the inset of Fig. \ref{fig_relThetaV_LamNat_LamApp}, showing the same relative variations for the slowness curve at $h=h_1=1.15\cdot10^{-21}$, $\eta=\eta_1=0.10253664614147$, but $\nu=0.95\cdot\nu_1=0.2995155288$. 

We have thus shown that in a sufficiently close vicinity of a higher-order caustic point, a broadband excitation in terms of wavenumber can result in a \emph{narrowband} CSWB with a very well-defined direction. As a result, this phenomenon is expected to be extremely favourable in experiments, since any realistic antenna cannot have an arbitrarily narrow excitation efficiency as a function of wavenumber. Provided that its design yields AC magnetic fields with Fourier components in the (broad) range of interest and with phases in a given interval of width $<\pi$, all the corresponding spin waves will coherently add in a beam with very small spectral breadth. In other words: in such a situation, counter-intuitively, exciting additional wave vectors with \emph{different} wavenumbers does not average out the carrier wave's amplitude but rather increase it. This naturally prompts the question of how much stronger the emission from a caustic point of higher order would be with respect to that of a regular caustic point, and more generally, of the spin wave amplitude enhancement due to the caustics. This, however, goes beyond the scope of the present manuscript.

To conclude this section, we point out that the reduced field  $h_\mathrm{c}(\eta)$ corresponding to the caustic point of higher order decreases as a function of reduced dipolar-exchange length. Thus, this feature is expected to exist only for $\eta<\eta_{\mathrm{c},0}\simeq 0.1025366$. 

\subsection{Merged caustic spin wave beams}

\begin{figure}%[!h]
\includegraphics[width=7.5cm]{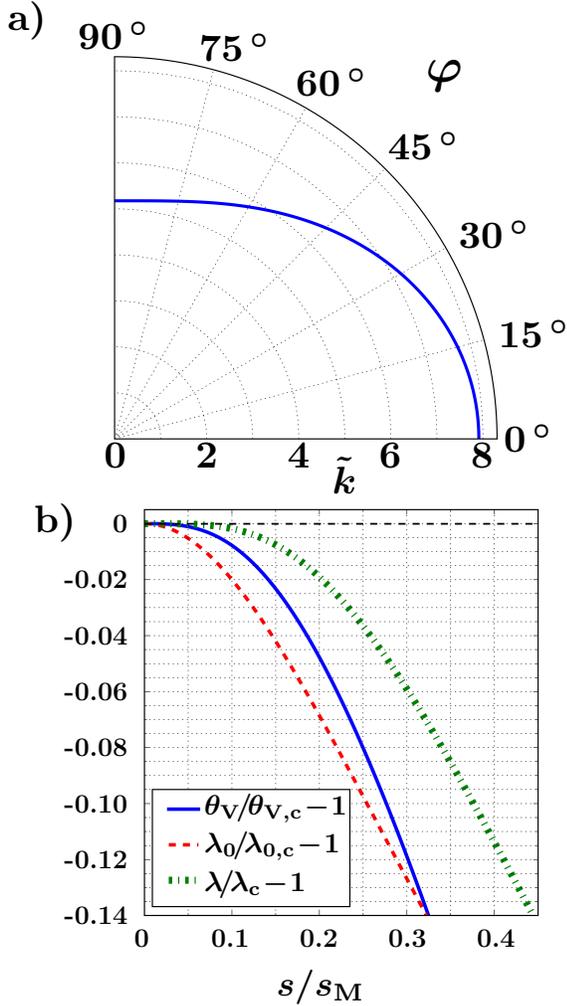}
\caption{\label{fig_MergedCSWB_RelVarAndSlownCurv}a) Slowness curve at ($\nu_2=0.71836419052,h_1,\eta_1$). b) Relative deviations in group velocity direction $\theta_\mathrm{V}$ (blue), natural wavelength $\lambda_0$ (red) and apparent wavelength $\lambda$ (green), as a function of curvilinear abscissa along the slowness curve normalized by its total length $s_\mathrm{M}$.}
\end{figure}

We now move on to the topic of the threshold frequency $\nu_\mathrm{m}(h,\eta)$ corresponding to the upper boundary of $\mathcal{D}$, \emph{i.e.} above which there are no caustic points any more. As was shown in Fig. \ref{fig_otherEtaVals_examplesThetaV}, the CSWB direction $\theta_\mathrm{V,c}$ goes to $\pi/2$ as $\nu\rightarrow\nu_\mathrm{m}(h,\eta)$. This is illustrated in Fig. \ref{fig_MergedCSWB_RelVarAndSlownCurv}, where we show a slowness curve for $\eta_1$, $h_1$, and $\nu_2=0.71836419052$. We stress again that $\nu_\mathrm{m}(h,\eta)$ is strictly speaking an infinitely narrow boundary and therefore $\nu_2\neq \nu_\mathrm{m}(h_1,\eta_1)$, but in these conditions, we find a unique caustic point on the slowness curve, with $\pi/2-\varphi_\mathrm{c}\simeq$\SI{0.32}{\micro\radian}, and $\theta_\mathrm{V,c}$ is equal to $\pi/2$ (within numerical precision). Moreover, at $\nu'_2=\nu_2+\delta\nu$, where $\delta\nu=1\cdot10^{-11}$, we do not find any caustic point on the slowness curve.

As a result, we take the slowness curve at ($\nu_2,h_1,\eta_1$) to be assimilable to the one at ($\nu_\mathrm{m}(h_1,\eta_1),h_1,\eta_1$). Its very straight aspect around $\varphi=\pi/2$ is somewhat reminiscent of the one seen in the discussion of the caustic point of higher order. To illustrate this in more detail, Fig. \ref{fig_MergedCSWB_RelVarAndSlownCurv}.b) displays the relative deviations in group velocity direction $\theta_\mathrm{V}$, natural wavelength and apparent wavelength around the caustic point at $\varphi_\mathrm{c}$. We point out that in the present case, the deviations are plotted against the curvilinear abscissa $s$ normalized to the slowness curve's length $s_\mathrm{M}$ instead of $\tilde{k}$ as in Fig. \ref{fig_relThetaV_LamNat_LamApp}. This choice is motivated by (i) the fact that in this case, to lowest order $\tilde{k}-\tilde{k}_\mathrm{c}=\mathcal{O}(s^2)$ instead of $\mathcal{O}(s-s_\mathrm{c})$ as before, and (ii) the much smaller relative difference between the smallest and largest normalized wavenumbers $\tilde{k}_\mathrm{m}$ respectively $\tilde{k}_\mathrm{M}$: $\tilde{k}_\mathrm{m}\simeq 5.17$ and $\tilde{k}_\mathrm{M}\simeq 7.91$, compared to $\tilde{k}_\mathrm{m}\simeq 0.240$ and $\tilde{k}_\mathrm{M}\simeq 5.66$ before. (i) implies that for ($\nu_2,h_1,\eta_1$), $\tilde{k}$ cannot serve as a meaningful abscissa along the curve since $\mathrm{d}\tilde{k}/\mathrm{d}s=0$, which was not the case for ($\nu_1,h_1,\eta_1$), while (ii) shows that the slowness curve for ($\nu_2,h_1,\eta_1$) is much closer to a fourth of a circle than that for ($\nu_1,h_1,\eta_1$); as a matter of fact, for ($\nu_2,h_1,\eta_1$), we find that $1-[\pi/2\cdot(\tilde{k}_\mathrm{m}+\tilde{k}_\mathrm{M})/2]/s_\mathrm{M}=3.7\%$. Therefore, $s/s_\mathrm{M}$ provides a better feeling for how much of the slowness curve contributes to the CSWB.

From the graph, it seems that the apparent wavelength has once more a quartic behaviour around the caustic point. We show in the Supplementary Materials that this is indeed the case: in the conditions where $\nu=\nu_\mathrm{m}(h,\eta)$, to the lowest non-zero order, $\theta_\mathrm{V}(s\rightarrow0)-\pi/2$ varies with an $s^3$ dependence around $s=0$, and the lowest-order variations in $\tilde{k}$ and $\varphi$ (around $\tilde{k}_\mathrm{m}$ and $\pi/2$) cancel each other out in the projection $\tilde{k}\cdot\cos{(\theta_\mathrm{V}-\varphi)}$.

As a result, a caustic point at $\nu_\mathrm{m}(h,\eta)$ is such that an excitation from a suitable, moderately directional antenna would be effectively narrowband, and weakly divergent around the group velocity direction $\theta_\mathrm{V,c}=\pi/2$. This orientation is itself also advantageous in practice: as long as the used antenna can excite sufficiently high wavenumbers, the CSWB direction becomes in this case simply perpendicular to the applied field. Moreover, owing to the symmetries of the dispersion relation, the CSWB benefits from the part of the slowness curve at $\varphi\gtrsim\pi/2$, which also feature $\theta_\mathrm{V}\simeq\pi/2$. That is why large spin wave amplitudes can be expected, as effectively two CSWB have merged at this particular frequency. We note that this merging phenomenon has already been observed in simulations by Kim \emph{et al.} \cite{Kim2016_PredictionOfSWcausticsInPerpFilmsWithDMI} in perpendicularly magnetized ultrathin films and by Gallardo \emph{et al.} \cite{Gallardo2021_SWcausticsInSynthAntiferromagnet} in synthetic antiferromagnets. For the sake of completeness, let us comment on what happens from an analytical point of view when the two caustic points just below and above $\varphi=\pi/2$ coincide. It must be kept in mind that they respectively correspond to a maximum and a minimum for $\theta_\mathrm{V}$, temporarily considered for $\varphi\in[0,\pi]$. Thus, when they do coincide at $\varphi_\mathrm{c}=\pi/2$, strictly speaking there is no caustic point any more. To put it differently: below $\nu_\mathrm{m}(h,\eta)$, over $\varphi\in [0,\pi]$, $\theta_\mathrm{V}$ increases up to the first caustic point where it reaches $\theta_\mathrm{V,c}>\pi/2$, decreases until the second one (for $\varphi>\pi/2$) where it reaches $\pi-\theta_\mathrm{V,c}$, then increases again to reach $\pi$ when $\varphi=\pi$. Exactly at $\nu_\mathrm{m}(h,\eta)$, it is monotonously increasing with an inflexion point, and above $\nu_\mathrm{m}(h,\eta)$, it is strictly monotonously increasing.

In order to go beyond the particular case presented here, we now investigate the evolution of $\nu_\mathrm{m}(h\rightarrow0,\eta)=\nu_{\mathrm{m},0}(\eta)$ as a function of reduced dipolar-exchange length $\eta$. Similarly to the caustic point of higher order, the evolutions as a function of reduced field quickly become cumbersome. This is why we focus on the $\nu_{\mathrm{m},0}(\eta)$, which is both the lowest frequency at which CSWBs merge and a threshold frequency that is easier to reach in experiments owing to the vanishing applied field, provided that the studied film is soft enough.

We do keep in mind that below a certain limit in terms of reduced dipolar-exchange length, the model we use loses its validity. However, it has been shown that at sufficiently high frequency \cite{Kreisel2009_FlawsOfsualSWdispersionInDipExchRegime}, the analytical dispersion relation derived by Kalinikos and Slavin describes spin waves once more with a good accuracy.

\begin{figure}[!h]
\includegraphics[width=8.595cm]{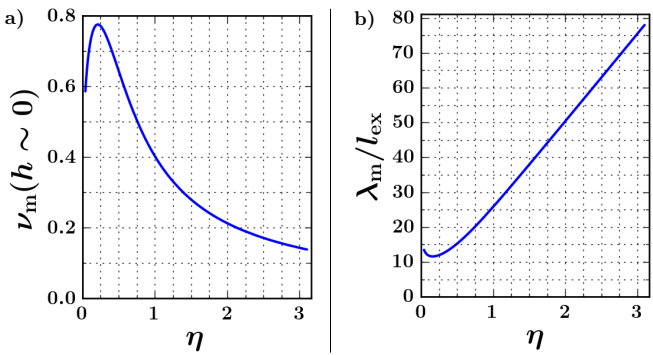}
\caption{\label{fig_NuMergeAndLambdaM_vs_eta}a) $\nu_{\mathrm{m},0}(\eta)$ as a function of reduced dipolar-exchange length $\eta$. b) The corresponding reduced wavelength $\tilde{\lambda}_m=(2\pi/\tilde{k})/\eta=\lambda/l_\mathrm{ex}=\lambda_0/l_\mathrm{ex}$. Here, natural and apparent wavelengths coincide as phase and group velocities are collinear.}
\end{figure}

Fig. \ref{fig_NuMergeAndLambdaM_vs_eta} displays the numerically determined dependence of $\nu_{\mathrm{m}0}$ on $\eta$, as well as that of $\lambda_\mathrm{m}/l_\mathrm{ex}$ the wavelength of the corresponding CSWB, normalized by the dipolar-exchange length. The procedure to find first a coarse estimate of this curve (before refining it with actual field and frequency maps) is described in the Supplementary Materials. We point out that in the case of the merged CSWBs, the apparent and natural wavelengths are equal since $\theta_\mathrm{V,c}=\varphi_\mathrm{c}=\pi/2$. The minimum value of $\eta$ in these graphs corresponds to the smallest one we used such that the slowness curve (in vanishing fields) has only one connected component. While we may not expect our findings to hold at the lowest $\eta$'s, we do expect their accuracy to improve as $\eta$ increases; it should be sufficient at least for $\eta>1$ since in this case the considered ferromagnetic film can truly be considered thin.

If we think about searching for the merged CSWBs, Fig. \ref{fig_NuMergeAndLambdaM_vs_eta}.b) indicates that for realistic values of $\eta=l_\mathrm{ex}/d$, the CSWBs' apparent wavelengths $\lambda$ are only about one order of magnitude larger than the material's dipolar-exchange length, typically $\lambda\lesssim 25 l_\mathrm{ex}$. This is in stark contrast with the case of the caustic point of higher order in vanishing field, where the \emph{natural} wavelength was $\lambda_{0,\mathrm{HO}}\simeq 84 l_\mathrm{ex}$, and the apparent wavelength $\lambda_\mathrm{HO} \simeq 334 l_\mathrm{ex}$. As a result, it seems that while caustic points of higher order may readily be excited by antennas created with even conventional electron beam lithography, in the case of the merged CSWB achieving a sufficient excitation efficiency at the proper wavevectors should prove quite challenging. For instance, even the low-magnetization, low-damping and rather soft ferrimagnet YIG features $l_\mathrm{ex}=$\SI{17.3}{\nano\meter} \cite{Klingler2014_fmrMeasurementOfYIGexchangeCst}, meaning that high-end antennas with a characteristic periodicity down to about \SI{200}{\nano\meter} would be required in this easiest of cases.

\section{Conclusions}

We have focused on some properties displayed by spin wave caustics in soft, thin ferromagnetic films. On the theoretical side, our approach relied on the analytical dispersion relation established by Kalinikos and Slavin. We could show that many reports on CSWBs in the literature can be interpreted within this frame, although the absence of characteristic signs of a true CSWB may still cause some ambiguity. Following up on most studies, we have performed time-resolved magneto-optical Kerr-effect-based microscopy on samples designed for the study of CSWBs. Despite the large thickness of the ferromagnetic material, our measurements are in very good agreement with our predictions, thus validating the approach. Furthermore, we have specifically highlighted the large misalignment between phase and group velocities in this case, and succeeded in observing narrow CSWBs. 

Just at the boundary of the dipolar-dominated regime accessible in our experiments, we have predicted the existence of a special caustic point. We refer to it as caustic point of higher order because it corresponds to an undulation point for the group velocity direction rather than a quadratic extremum. This configuration was shown to be of particular interest because the apparent wavelength also featured a quartic behaviour, which implies a low spectral breadth for the CSWB even in the case of a broadband excitation. Although we focused on the special value $\eta_\mathrm{c}$ of reduced dipolar-exchange length such that the caustic point of higher order occurs at vanishing applied fields, we stress that this phenomenon would appear at non-zero fields for $\eta<\eta_\mathrm{c}$, as long as the dispersion relation we use is valid.

Finally, we have investigated the merging of CSWBs. Once again, we have studied in detail the case of vanishing applied fields, yet the merging may occur for any field value, provided that the excitation frequency is large enough. In terms of model validity, it must be recalled that while vanishing values of $\eta$ are problematic for the chosen dispersion relation, the merging always occurs at frequencies close to the exchange-dominated regime. The discrepancies between the actual spin wave dispersion and the model by Kalinikos and Slavin decrease in this frequency range \cite{Harms2022_DipoleExchangeSpinWaveTheory}. As a result, our claim is that the merging frequencies $\nu_\mathrm{m0}$ obtained for low $\eta$ may be slightly inaccurate yet the phenomenology should remain the same as for larger $\eta$, where we expect our predictions to be more reliable. As the CSWBs merge, a very significant portion of the slowness curve contributes to spin wave emission around $\theta_\mathrm{V}=\varphi=\pi/2$. Therefore, this configuration appears promising in terms of channelling strong spin wave beams with short wavelengths, as low as $\sim 15 l_\mathrm{ex}$.

One of the most important questions remaining unaddressed so far concerns the quantification and prediction of the enhancement of amplitude associated with CSWBs. More precisely, the crucial distinction between natural and apparent wavelength as well as the inadequacy of the usual Huygens-Fresnel approach (due to the strong non-collinearity between phase and group velocities) in the construction of CSWBs calls for alternative evaluations of their amplitudes. We intend to clarify these points and to go beyond the usually described amplitude divergence so as to reconcile the theoretically vanishing curvature (on the slowness curve) and experimentally finite amplitudes.

%\appendix
%\section{}

%\begin{acknowledgments}
%\end{acknowledgments}

\bibliography{bibliography_caustics.bib}

\end{document}